\documentclass[aps,prl,twocolumn,showpacs,groupedaddress]{revtex4-1}
\usepackage{}
\usepackage{pifont}
\usepackage{amsmath}
\usepackage{graphicx}
\usepackage{amssymb}
\usepackage{amsfonts}
\usepackage{subfigure}
\usepackage{booktabs}
\usepackage{setspace}
\usepackage{threeparttable}
\usepackage{enumerate}

\begin{document}

\title{Itinerant magnetism in Weyl spin-orbit coupled Fermi gas}
\author{Shang-Shun Zhang$^{1,2}$}
\author{Jinwu Ye$^{2,3}$}
\author{Wu-Ming Liu$^1$}
\affiliation{ $^{1}$Beijing National Laboratory for Condensed Matter Physics, Institute of
Physics, Chinese Academy of Sciences, Beijing 100190, China \\
$^{2}$ Department of Physics and Astronomy, Mississippi State University, MS 39762, USA  \\
$^{3}$ Key Laboratory of Terahertz Optoelectronics, Ministry of Education, Department of Physics, Capital Normal University, Beijing 100048, China
}
\date{\today}

\begin{abstract}
\textbf{Magnetic ordering of itinerant fermionic systems is at the forefront of condensed matter physics dating back to Stoner's instability. Spin-orbit coupling (SOC) which couples two essential ingredients of an itinerant fermionic system, namely spin and orbital motion, opens up new horizons to this long-standing problem. Here we report that the itinerant ferromagnetism is absent in 3D Fermi gas with a Weyl SOC and various itinerant spin density waves emerge instead, which is deeply rooted in the unique symmetry and spin-momentum locking effect in spin-orbit coupled systems. What is more appealing is that, the strong SOC provides a new and efficient mechanism to realize itinerant spin density waves at extremely weak repulsion---a significant benefit for present ultra-cold atom experiment. These novel phenomena can be probed by Bragg spectroscopy, time of flight imaging and {\sl In-Situ } measurements in ultra-cold atom experiment.}
\end{abstract}


\maketitle

Itinerant magnetism in fermionic systems dating back to Stoner's instability \cite{stone,millis1,millis2} is a long-standing concerned focus
in material science, in which magnetic particles such as d electrons in transition metals can move freely in the entire object.
It is thought that the itinerant magnetism is a result of the competition between the repulsive interaction and Pauli exclusion principle, which is known as the Stoner's mechanism \cite{stoner1,stoner2,stoner3,stoner4}.
However, this scenario is far from conclusive in consequence of the strong coupling nature of this issue where various correlation effects beyond the Stoner's mechanism set in \cite{corre1,corre3,corre4,corre5,corre7}.
Recently, ultra-cold atom has shed light on the study of itinerant magnetism, by which simple systems with highly tunability and purity can be created. To date, there have been both experimental \cite{GBJo} and theoretical work \cite{fmcold11,fmcold14,fmcold15,fmcold16} on possible itinerant magnetism in the context of cold atoms, including purely repulsively interacting two component Fermi gases \cite{stoner2,stoner3,stoner4,corre5,two1,two2}, repulsive polarons in strongly polarized Fermi gases \cite{polaron2,polaron3} and multi-orbital systems on optical lattices \cite{pband1}.
However, the itinerant magnetism in fermionic systems has not yet been observed in recent experiments.

Besides, the investigation and control of the spin-orbit coupling (SOC) in ultra-cold atoms has become an exciting subject over recent years \cite{revSOC1,revSOC2,revSOC4}. Thanks to the high precision control of light-matter interaction, Abelian or non-Abelian gauge fields have been engineered in both degenerate Bose and Fermi gas \cite{SOCexp1,SOCexp2,SOCexp3}, which inspired extensive theoretical investigations on various important effects of SOC on the pairing physics of attractively interacting Fermi gases
\cite{soctheory31,soctheory32,soctheory33,soctheory34,soctheory35,soctheory36,soctheory37,soctheory38,pairing}.
Even so, dramatic effects available induced by SOC on the itinerant magnetism in fermionic systems have not been discussed so far.
According to our observation, SOC is absolutely nontrivial to itinerant magnetism: on the one hand, it essentially couples the magnetism to orbital motion and dramatically changes the symmetry of systems without SOC; on the other hand, it strongly modifies the Fermi surface structure which supports different types of particle-hole excitations from intra- or inter- Fermi surfaces and greatly enhances single-particle density of states which creates a favorable condition to realize itinerant magnetism at an extremely weak interaction strength.
Given these huge differences between systems with and without SOC, it is naturally expected that there would be rich and amazing physics unexplored previously on itinerant magnetism of spin-orbit coupled fermionic systems \cite{CJWu0,SSZhang,topo,rhboson}.

In this work, we have systematically addressed this problem by investigating rich competing magnetic orderings in spin-orbit coupled fermionic systems. Remarkably, we find that itinerant ferromagnetism is absent in three dimensional Fermi gas with a Weyl SOC but instead various itinerant spin density waves emerge.
The symmetry group of spin-orbit coupled system which is very
different from any previously known (non)-relativistic systems plays a critical role in these huge differences than systems without SOC.
In the paramagnetic respect, we identify one gapless sound mode and three gapped modes. Due to the lack of inversion symmetry in spin-orbit coupled system , two
transverse modes corresponding to collective spiral spin fluctuations split at any finite momenta ($\omega_{T_{\pm}}=\Delta\pm \beta q$), which indicates a possible spiral transverse spin density wave (TSDW) instability
of this system driven by collective modes.
From the exact constraints by symmetry, we show that a putative ferromagnetism is always unstable against a possible TSDW phase.
Indeed, as interaction increases, the paramagnetic state becomes a spiral TSDW phase at small or intermediate SOC
or a stripe longitudinal (collinear) spin density wave (LSDW) at large SOC. The transition from TSDW to LSDW phase is attributed to the spin-momentum locking effect which leads to anisotropy in spin space.
Another remarkable property of magnetic orderings in spin-orbit coupled systems is that the critical interaction strength is dramatically reduced for strong SOC due to the nearly flat band structure near the bottom (Weyl shell) of the spectrum.
As a result, all these novel phenomena could be observed at a relatively weak interaction strength, which is a significant benefit for cold-atom experiment.
Viewing the rapid technical developments of cold atoms in Bragg spectroscopy \cite{bragg1,bragg3,bragg6,light1,light2},
speckling and {\sl In-Situ} measurements \cite{speckle1,speckle2,Shin,Gemelke}, our theoretical predictions could be probed in near future experiments in both cold atoms and relevant condensed matter systems.

\bigskip

\textbf{\large{Results}}

\textbf{The system and symmetry analysis of the Hamiltonian}
We consider a repulsively interacting two-component Fermi gas with isotropic Weyl SOC $ V_{SO}= \lambda \vec{k} \cdot \vec{\sigma} $ described by the Hamiltonian:
\begin{eqnarray}  \label{model2D}
 H & = & \int d^{3} \vec{r} \Psi^{\dagger} ( \frac{ -\hbar^{2} \nabla^{2}}{ 2m}-\mu + V_{SO} ) \Psi
                             \nonumber   \\
    & + & g \int d^{3} \vec{r}  \Psi^{\dagger}_{\uparrow}(  \vec{r} )\Psi^{\dagger}_{\downarrow}(  \vec{r} )
     \Psi_{\downarrow}(  \vec{r} )\Psi_{\uparrow}(  \vec{r} ),
\label{weyl}
\end{eqnarray}
where $\lambda$ refers to the strength of SOC which has been proposed to be simulated in cold atoms by several schemes \cite{3DSOC1,3DSOC2,3DSOC3} and $g=4\pi \hbar^2 a_s/m$ with $a_s$ the $s$-wave scattering length. The chemical potential $\mu$ is self-consistently determined by the density of atoms.
In spin-orbit coupled systems, it is convenient to take $k_R = m\lambda$ as the momentum unit and $E_R=k_R^2/2m$ as the energy unit.
The effect of SOC term is characterized by the dimensionless ratio $\gamma=k_R/k_F$ where
$k_F=(6\pi^2n)^{1/3}$ is the Fermi momentum of the system without SOC at the same density $ n $. Hereafter,
we set $\hbar=k_B=1 $.

The symmetry analysis for interacting  spin-orbit coupled systems
is very different from any previously known (non)-relativistic systems and have never been systematically discussed before
\cite{pairing,rhboson}.
Ahead of performing any analytical or numerical calculations, it is important to investigate symmetries of systems and
their implications on experimentally observable physical quantities. Here, we classify the symmetries of Weyl SOC Hamiltonian Eq. (\ref{weyl}) and find their exact constraints on various density-spin correlation functions. The symmetries of Hamiltonian Eq. (\ref{weyl}) are: i) continuous $[SU(2)_{spin} \times SO(3)_{orbit} ]_D $, where the subscript $ D $ represents the simultaneous rotations in both spin and orbital space;
ii) the time reversal symmetry $\mathcal{T}$: $\mathcal{T}\Psi_{\vec{k}\uparrow}\mathcal{T}^{-1}=\Psi_{-\vec{k}\downarrow}$ and $\mathcal{T}\Psi_{\vec{k}\downarrow}\mathcal{T}^{-1}=-\Psi_{-\vec{k}\uparrow}$, which implies $ \vec{s} \rightarrow -\vec{s}, \vec{k} \rightarrow -\vec{k}, i \rightarrow -i $;
iii) three spin-orbital coupled discrete symmetries: $\mathcal{P}_x=[\mathcal{I}_x\times\mathcal{S}_x]_D$ where $\mathcal{I}_x: p_x\rightarrow p_x,p_{y,z}\rightarrow -p_{y,z}$ is the spatial reflection operator about $x$ axes and $\mathcal{S}_x: s_x\rightarrow s_x,s_{y,z}\rightarrow -s_{y,z}$ is a $\pi$ rotation about $s_x$ axes, $\mathcal{P}_y=[\mathcal{I}_y\times\mathcal{S}_y]_D$ and $\mathcal{P}_z=[\mathcal{I}_z\times\mathcal{S}_z]_D$. Note that the $\mathcal{P}_z$ is just a special case of  $[SU(2)_{spin} \times SO(3)_{orbit} ]_D $ symmetry with a $ \pi $  rotation about $s_z$ axes. Obviously, all of the symmetry operations listed above act on both spin and orbital freedoms. These symmetries can be used to establish exact relations on the spin-density correlation functions, some of which are listed here and the details are given in the supplementary Material.

\begin{figure}[!t]
\begin{center}
\includegraphics[width=3.3in]{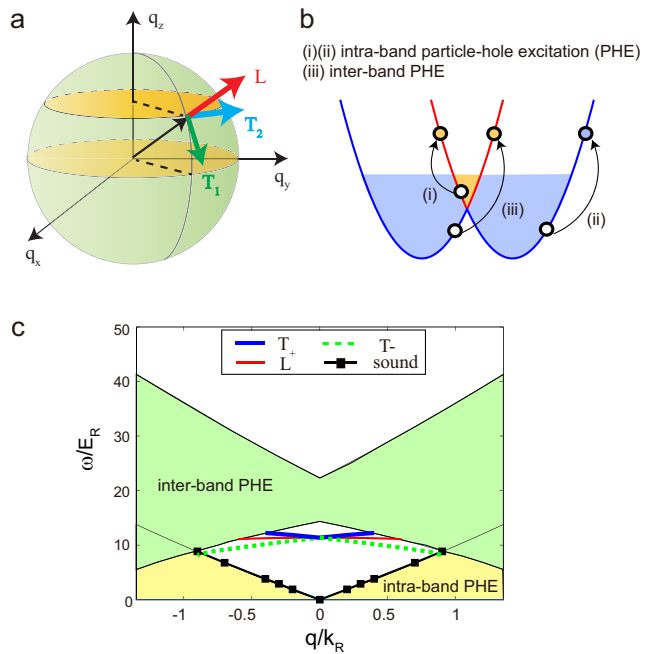}
\end{center}
\caption{ \textbf{Helical spin basis, particle-hole excitation and the collective modes.}
(\textbf{a}) In systems with Weyl SOC, it is convenient to define the helical bases shown here: one longitudinal spin (L defined by $\hat{L}= \hat{q}$) and two transverse spin ($T_1,T_2$ defined by $\hat{T}_{1,2}\cdot \vec{q}=0$).
(\textbf{b}) Schematic plots of the intra-band (process (i) and (ii)) and inter-band (process (iii)) particle-hole excitations, which are the basic quasi-particle excitations and are responsible for the decay of the collective modes shown in \textbf{c}.
(\textbf{c}) The collective modes in the normal state with SOC strength $\gamma=0.28 $: the sound mode and the L mode are combined oscillations of density and $\hat{L}$ spin fluctuations shown in \textbf{a}, the two transverse modes $T_+$/$T_-$ are right/left-handed spiral spin fluctuations formed by $T_{\pm}=(\hat{T}_1\pm\hat{T}_2)/\sqrt{2}$ with $\hat{T}_{1,2}$ the transverse spin fluctuations shown in \textbf{a}.
The yellow (green) regime represents the continuum of intra-(inter-) band particle-hole excitations shown in \textbf{b}. All these collective modes are
damped when entering the continuum through the proliferations of particle-hole excitations.}
\label{fig:collA}
\end{figure}

In the presence of SOC, it is convenient to define a helical bases (Fig. 1 \textbf{a})
where we have introduced two transverse spin components ($\hat{T_1}\cdot \vec{S}$, $\hat{T_2}\cdot \vec{S}$) and one
longitudinal spin component ($\hat{L}=\hat{q}\cdot \vec{S}$).
The $[SU(2)_{spin} \times SO(3)_{orbit} ]_D $ symmetry dictates
 that dynamic   $ 4 \times 4 $ density-spin response function $ \chi^{\mu\nu}(\vec{q}, \omega ) $ can be split into two $ 2 \times 2 $ subspaces:
 i) $n$(density)-$L$ subspace; ii) $T_1$-$T_2$ subspace. Due to the combined rotational symmetry, one can always pick up one direction to present all of the physics, say, $\hat{z}$ direction as we have done in the following sections. To illustrate the relationship between $\pm \vec{q}$ using the discrete symmetries $\mathcal{P}_{x,y}$ and $\mathcal{T}$, we will use a unified helical basis for $\pm \vec{q}$ as done here for symmetry argument.

  In the  $T_1$-$T_2$ subspace, the combination of $\mathcal{P}_x$(or $\mathcal{P}_y$) symmetry and time reversal symmetry $\mathcal{T}$ implies that:
\begin{equation}
 \chi(\vec{q}, \omega)=\begin{pmatrix}
		\chi^{+-}(\vec{q}, \omega) & 0 \\
		0 & \chi^{-+} (\vec{q}, \omega)\\
	\end{pmatrix},~~\chi^{+-}(\vec{q}, \omega)=  \chi^{-+}(-\vec{q}, \omega),
\label{pm}
\end{equation}
which is on the basis of $s^{\pm}=s^x\pm i s^y$ bases via a unitary transformation. Besides, there is another conjugate relationship: $[\chi^{+-}(\vec{q},\omega)]^*=\chi^{-+}(-\vec{q},-\omega)$. The poles of $\chi^{+-}$ and $\chi^{-+}$ lead to two gapped transverse collective modes $ \hat{T}_{\pm}={1\over \sqrt{2}}(\hat{T}_1\pm i \hat{T}_2)$ respectively. The $\hat{T}_{\pm}$ modes are collective oscillations of right/left-handed chiral spin fluctuations: $\vec{\phi}_T=\phi_0(\cos(qz-\omega_q t),\pm \sin (qz-\omega_qt),0)$ (here we have assume $\vec{q}=q\hat{e}_z$). In the $n$-$L$ subspace, the application of $\mathcal{P}_x$(or $\mathcal{P}_y$) symmetry and time reversal symmetry $\mathcal{T}$ leads to
\begin{eqnarray}\label{eq:nl}
 \chi^{nn}(\vec{q}, \omega) & = &  \chi^{nn}(-\vec{q}, \omega), \chi^{LL}(\vec{q}, \omega)=  \chi^{LL}(-\vec{q}, \omega)
     \nonumber  \\
 \chi^{nL}(\vec{q}, \omega) & = & \chi^{Ln}(\vec{q}, \omega) = - \chi^{nL}(-\vec{q}, \omega).
\label{pxnl}
\end{eqnarray}
There is also a conjugate relationship: $[\chi^{ij}(\vec{q},\omega)]^*=\chi^{ij}(-\vec{q},-\omega)$. The poles of the susceptibility in the $ n$-$L $ space give birth to a gapless sound mode and a gapped $ L $ mode, both of which are coupled oscillations of density and longitudinal spin.

These exact relations above cause rigorous constraints on the dispersions of the collective modes of coupled density-spin fluctuations and the form of effective action in the low energy particle-hole channels. Moreover, they indicate that the magnetic instabilities in the transverse and longitudinal spin channels are independent, which form two basic putative competing channels with details discussed bellow. Our path-integral formulism applied here is found to respect all these exact relations.

\begin{figure}[!t]
\begin{center}
\includegraphics[width=3.4in]{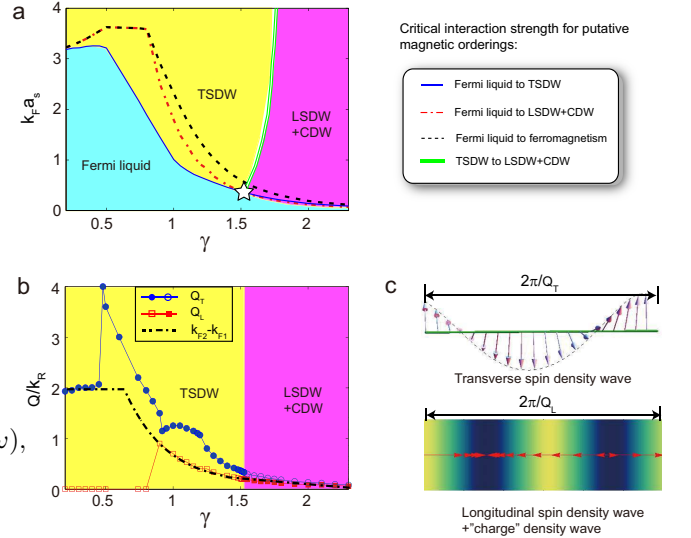}
\end{center}
\caption{\textbf{Phase diagram and the orbital ordering wavevector.}
(\textbf{a}) The critical interaction strength from the Fermi liquid (paramagnetic state) to itinerant TSDW (blue line),
LSDW+CDW (red dashed dotted line) or a putative ferromagnetism (black dashed line) transition.
The three curves show the competing orders among the 4 phases.
The itinerant ferromagnetism is always unstable against the TSDW, in sharp contrast to a three dimensional Fermi system without SOC.
The transition from the Fermi liquid to the TSDW or LSDW+CDW is of the bosonic Lifshitz type, that from TSDW to LSDW+CDW is first order.
(\textbf{b}) The ordering wavevector $Q_T$ (blue circles) and $ Q_L $ (red squares) of the TSDW and LSDW as a function of SOC strength $\gamma$.
The green dot-dashed line represents the $k_{F2}-k_{F1}$.
When $ \gamma < 0.5 $, $ Q_T \sim k_{F2}-k_{F1} $ is dominated by the momentum transfer between outer Fermi surface and inner Fermi surface with the opposite helicity,
When $ 0.5 < \gamma < 1.5 $, $ Q_T $ is dominated by other momentum transfer processes. When
$ \gamma > 1.5 $, $ Q_L \sim k_{F2}-k_{F1} $ is dominated by the momentum transfer between outer Fermi surface and inner Fermi surface with the same helicity.
(\textbf{c}) Schematic plots of the TSDW and LSDW+CDW orders, where the arrow indicates the direction of spins, the background with varying colors shows the CDW background induced by the LSDW orders, whose spatial period is half of that of the LSDW order.}
\label{fig:mf3D}
\end{figure}

\textbf{Collective modes in the paramagnetic state}
Within the path-integral framework developed in the Methods, one can extract collective modes from the poles of density-spin response functions at the random phase approximation (RPA) level.
In the $T_1$-$T_2$ subspace, there are two split spiral transverse modes
$ \omega_{T_{+/-}}(\vec{q})=\Delta \pm \beta q  $ (Fig. 1 \textbf{c}).
The $ T_{+} $ mode is the pole of $ \chi^{+-} $, while the $ T_{-} $ mode is the pole of $\chi^{-+} $.
The split of the two transverse modes for finite $ q $ is due to the absence of inversion symmetry (or equivalently $\mathcal{S}_x$ or $\mathcal{S}_y$ symmetry) of the Weyl SOC term (see Supplementary Material for details).
In the $n$-$L$ subspace,
the mixing of density and longitudinal spin fluctuation leads to one gapless sound mode $\omega^s_{\vec{q}}=v_s q$
and one gapped (called $ L $) mode $\omega_L(\vec{q})=\Delta+\alpha q^2$ (Fig. 1 \textbf{c}).
The dispersions of all the three gapped modes $T_{+}$, $T_{-}$ and $L$ can be straightforwardly deduced by the pole equation $\det(\chi^{-1})=0$ combined with
the symmetry constraints established in Eq. (\ref{pm}) and (\ref{eq:nl}).
The restored $SU(2)_{spin}$ symmetry at $\vec{q}=0$ dictates that the three gapped modes must have the same gap $ \Delta $ at $ \vec{q}=0$.
We find that $\Delta$ vanishes for interaction strength larger than $g_s=24\pi^2k_R/m(k_{F2}^2-k_{F1}^2)$, where $k_{F2}>k_{F1}$ are the two Fermi momenta in the presence of SOC. This corresponds to an instability driven by the collective modes at $ \vec{q}=0 $. Both the symmetry constraints and RPA calculations (Fig. 1 \textbf{c}) indicate that $\omega_{T_{-}}(\vec{q})=\Delta - \beta |\vec{q}|$ which may become negative at $|\vec{q}|= Q_T $ before $ \Delta=0 $ at $ \vec{q}=0 $. This indicate a transverse (spiral, chiral and co-planar) SDW transition at $ Q_T \hat{z}$ with the
order parameter $ \vec{\phi}_{T}= \phi_0 ( \cos Q z, -\sin Q z, 0 ) $ (here, we have pick up the $z$ direction).
Such an instability and its competitions with various other possible instabilities will be discussed next.

\textbf{Absence of ferromagnetic instability}
There is a ferromagnetic (FM) instability when the static spin susceptibility diverges at $ \delta={1\over g_c}-{\chi_{0}\over 4}=0 $, i.e.,
the Stoner FM instability. Here $\chi_0( \vec{q} \rightarrow 0, \omega=0)=\frac{m}{6\pi^2}(\frac{k_{F1}^2+k_{F2}^2}{K}+\frac{k_{F2}^2-k_{F1}^2}{k_R})$ with $ K=\sqrt{k_R^2+2m\mu} $ is the static spin susceptibility of 3D non-interacting Fermi gas with the Weyl SOC.
Then, we can obtain the critical value of dimensionless interaction strength $k_F a^{c}_s=F(\gamma)$ as shown in Fig. 2 \textbf{a}.
As $ \gamma $ increases, $F(\gamma)$ has a maximum at $\gamma \simeq 0.63$ corresponding to the chemical potential $\mu=-{2\over 3} E_R$,
after which, $k_F a^c_s$ decreases quickly. With a strong SOC, the $\mu$ approaches the bottom of the spectrum (the Weyl shell at  $|\vec{k}|=k_R$) where the density of states diverges as ${1\over\sqrt{\epsilon}}$. The effect of an interaction is dramatically enhanced, even a weak interaction may drive the system into the itinerant FM state.

However, this FM instability is not the leading instability of this system, although it happens before that driven by the $\vec{q}=0$ instability of the collective modes as we find $g_c$ is always smaller than $g_s$. In order to figure out this problem, it is tempting to construct
a quantum Ginburg-Landau action to describe the putative paramagnet-ferromagnetism transition, in which the spatial spin fluctuation is captured.
Taking the $\vec{q}\rightarrow 0, \omega/v_F q \rightarrow 0$ limit and integrating out the non-critical density mode,
one can obtain the effective action in terms of the spin fluctuation order parameters consistent with all the symmetries of Hamiltonian:
\begin{eqnarray}\label{eq:Hertz1d1}
\mathcal{S}  &=& \!\!\! \int \!\!\! \frac{d^3\vec{q}}{(2\pi)^3}T\sum_{n} \frac{1}{2}\mathcal{G}^{-1}_s \hat{P}_s^{ij}\phi_i^{*}\phi_j(\vec{q},i\omega_n)+u \!\! \int \!\! d^3\vec{r}d\tau (\vec{\phi}^2)^2,
\label{actionfm}
\end{eqnarray}
where $\hat{P}_s^{ij}=\hat{n}_s^{i*} \hat{n}_s^j$ ($\hat{n}_s=\hat{T}_+, \hat{T}_-, \hat{L}$ with $\hat{T}_{\pm}={1\over \sqrt{2}} (\hat{T}_1\pm i \hat{T}_2)$ shown schematically in Fig. 1 \textbf{a}) is the projection operator into helical bases, $\mathcal{G}_{T_{\pm}}^{-1}=\delta+ \gamma_T | y | \pm \beta_T q + \alpha_T q^2  $ and
$\mathcal{G}_L^{-1}=\delta+\gamma_L y^2+\alpha_L q^2$ with $ y=\omega_n/v_F q $ are propagators of the helical spin modes.
We observe that the $\pm q$ terms in the transverse propagators are dictated by the exact symmetry Eq. (\ref{pm}), which indicates that the putative FM state is always unstable towards a TSDW phase at a finite $q$. This conclusion is
reached by foregoing exact spin-orbital symmetry analysis, therefore independent of any approximations.
It is worth noting that the TSDW instability is driven not by the drop of the collective modes at a finite $\vec{q}$ but by the proliferation of low energy left-handed spiral spin fluctuations with finite $q$ described by the effective action in Eq. (\ref{eq:Hertz1d1}), which is supported by numerically monitoring the behavior of collective modes across the phase transitions.

\textbf{ Transverse spin density wave at smaller spin-orbit coupling ($ \gamma < 1.5 $)}
The exact symmetry analysis discussed above points to the formation of TSDW order before FM order and also possible orders driven by collective modes.
Here we search for the pole of $ \chi^{-+}(Q_T, \omega=0)$ to find out the condition of the appearance of TSDW phase, where the static transverse spin
susceptibility diverges. Numerical results of $\chi^{-+}(Q_T, \omega=0)$ can be found in the Methods.
The critical interaction strength $ k_F a_s $ and the orbital momentum $ Q_T $ are shown in Fig. 2 \textbf{a} and 2 \textbf{b} respectively. The critical interaction strength corresponding to the putative FM state is also shown there, which is found to be always larger than that of the TSDW phase. The order parameter is given by $\vec{\phi}_T=\phi_0(\cos (Q_T z), -\sin (Q_T z),0)$ (assuming $\vec{Q}_T=Q_T \hat{z}$), which is a spiral SDW order. Note that this order is left-handed, and its helicity is directly related to the sign of SOC term. Interestingly, any local magnetic orders could induce a density current in spin-orbit coupled systems. Here, we find the TSDW order would induce a spiral density current given by $\vec{j}=\lambda \vec{\phi}_T$, which does not exist in the paramagnetic state or magnetic state of systems without SOC. Therefore, it provides a silent feature of the presence of TSDW order and can be detected easily through time of flight measurement in cold atom experiment.

It is interesting to note that there are various jumps of the orbital wavevector $Q_T$ as increasing SOC strength (see Fig. 2 \textbf{b}). Physically, the jump of $Q_T$ is in consequence of the multiple maximums of the spin susceptibility and their competition with increasing SOC (see Fig. 5). As schematically shown by Fig. 1 \textbf{b}, the Fermi gas with SOC has two distinct Fermi surfaces, which support different types of low-energy particle-hole excitations with momenta $Q=2k_{F1},2k_{F2},k_{F1}+k_{F2},k_{F2}-k_{F1}$ (Fig. 1 \textbf{b}). Usually, the spin susceptibility is enhanced at these momenta, for example, we have identified the susceptibility is dominated by $k_{F2}-k_{F1}$ at small and large SOC strength whose mechanism has been illustrated in the following sections (also shown by Fig. 5 \textbf{a},\textbf{d}). In between, the maximum of the susceptibility is not dominated by a single momentum as listed above (see Fig. 5 \textbf{b},\textbf{c}). Nevertheless, we still find multiple maximums at different momenta and compete with each other as increasing SOC. This phenomenon is closely related to the complicated Fermi surface structure of spin-orbit coupled systems and brings forth interesting experimental signatures.

Then, within the path-integral formulism, one can construct an effective action to describe the itinerant paramagnet (PM) to the TSDW transition.
After integrating out the massive $ \phi_L $ mode, $\phi_{T_+}$ mode and the density mode,
we find the effective action takes a similar form as Eq. (\ref{actionfm}) where $ \vec{\phi} \rightarrow \phi_{-}$
and the transverse propagator is given by
$\mathcal{G}_{T_{-}}^{-1}(\omega_n,\vec{q})=\delta_T + \gamma_T | \omega_n | + \alpha_T (|\vec{q}| - Q_T)^2 $ with the dynamic exponent $ z_T=2 $.
When $ \delta_T > 0 $, the system stays in the PM state where $ \langle \phi_{-} \rangle = 0 $.
When $ \delta_T < 0 $, the system stays in the stripe TSDW phase where $ \langle \phi_{-} \rangle = \phi_0 e^{ i ( Q_T z) } $.
This is a bosonic Lifshitz transition which exists in various condensed matter systems such as the superfluid $^{4}$He \cite{roton01,roton02},
exciton superfluids in bilayer quantum Hall or electron-hole bilayer \cite{roton11,roton12}
and superconductor in a Zeeman field \cite{LO1,LO2}.
It is a first order phase transition leading to the stripe \cite{LO1,LO2} TSDW phase.

Finally, let's consider the low energy excitations above the TSDW phase. The symmetry breaking from PM to TSDW phase is $ [SU(2)_{spin} \times SO(3)_{orbit} ]_D \times Tran \rightarrow [U(1)^{\alpha}_{spin} \times U(1)^{\alpha}_{orbit} \times ( z \rightarrow z+ \alpha/Q_T ) ]_D $ with $Tran$ referring to a continuous translation and $\alpha$ the angle of the combined rotation. Of course, when $ \alpha= 2 \pi $, it reduces to just a translation by a lattice constant $ z \rightarrow z+ a $ with $a=2\pi/Q_T$. This symmetry breaking leads to one gapless Goldstone mode $ \theta $ in
$ \vec{\phi}_T= \phi_0 (\cos ( Q_T z + \theta ), -\sin ( Q_T z+ \theta ),0 ) $ which is a coupled lattice phonon and transverse spin mode.
By a symmetry analysis, we obtain the effective action of the gapless Goldstone mode
 $ {\cal L}_T[\theta ] =( \partial_{\tau} \theta )^2 + a_T ( \partial_z \theta )^2  +  b_T ( \partial_{\perp} \theta )^2 $.
There is a finite temperature 3D XY transition driven by the vortex un-binding in the phase of
the mixed Goldstone mode $ \theta $ where the TSDW phase also melts.

\textbf{Longitudinal spin density wave at larger spin-orbit coupling ($ \gamma > 1.5 $)}
Constrained by the spin-orbital symmetry, the susceptibility corresponding to longitudinal spin channel is decoupled to that of transverse spin channels and it provides another putative competing order, i.e., longitudinal spin density wave (LSDW). Taking similar procedure as above, we obtain the critical interaction strength from paramagnet to LSDW phase and the orbital wavevector $Q_L$ as shown in Fig. 2. By comparing the critical interaction strengthes in different channels, we find that the TSDW phase turns into LSDW phase $ \vec{\phi}_L= \phi_0 ( 0 ,0, \cos ( Q_L z)  ) $ with the ordering wavevector $ Q_L \sim k_{F2}-k_{F1} $ when the filling of fermions approaches to the bottom of the Weyl shell $ \gamma > 1.5 $.
Similarly, the local LSDW order induces a density current $\vec{j}=\lambda \vec{\phi}_L$ which also brings forth detectable experimental signatures. Besides, there is an accompanying ``charge'' density wave order (CDW) with the density ordering wavevector $ Q_C=2 Q_L $ by contrast to the TSDW case: $ \delta \phi_n=  \phi^2_0 \cos ( 2 Q_L z) $.
Note that at the static limit $ \omega=0 $, $\chi_{ij}$ with $(i.j)\in(n,L)$ is Hermite and therefore $ \chi_{ij} (\vec{q},0)$ is real based on Eq. (\ref{pxnl}). The real part of $\chi_{nL}(\vec{q},\omega)$ is an odd function of $\omega$ which implies $ \chi_{nL} (\vec{q},0) =0 $, namely there is no coupling between the density and longitudinal mode at the quadratic level.
We also find $ \chi_{nn} (Q_L, 0 ) $ is kept finite as approaching to the critical interaction strength, so the transition is {\sl not} charge driven.
Following the classical Ginzburg-Landau action at $ \omega=0 $ constructed in \cite{ek1,ek2}, we can see that the CDW is driven by the spin-ordering due to the cubic coupling $ \lambda_3 \delta \phi_n \phi^2_L $.
After substituting the parasitic CDW order \cite{ek1,ek2} and integrating out the two massive transverse modes $  T_{\pm} $, we find the effective action describing the PM to LSDW+CDW phase transition takes a similar form as Eq. \ref{actionfm} where $ \vec{\phi} \rightarrow \phi_L $ and the longitudinal propagator is $\mathcal{G}_L^{-1}=\delta_L+\gamma_L |\omega_n|+ \alpha_L (|\vec{q}|-Q_L)^2 $ with the dynamic exponent $ z_L=2 $.
When $ \delta_L > 0 $, the system stays in the PM state where $ \langle \phi_L \rangle = 0 $.
When $ \delta_L < 0 $, the system stays in the stripe LSDW phase where $ \langle \phi_L \rangle = \phi_0 e^{i Q_L z} $.
This is also a bosonic Lifshitz type of phase transition \cite{roton01,roton02,roton11,roton12,LO1,LO2}.

The symmetry breaking from the PM to the LSDW + CDW is $[SU(2)_{spin} \times SO(3)_{orbit} ]_D \times Tran
\rightarrow [U(1)^z_{spin} \times U(1)_{orbit} ]_D \times ( z \rightarrow z+ 2 \pi/Q_L )$. It leads to one gapless lattice phonon mode $ \theta $ in
$\vec{\phi}_L= \phi_0 ( 0 ,0, \cos ( Q_L z+ \theta))$.
By symmetry analysis and drawing the analogy from the smectic liquid crystal \cite{book:Lubensky}, we get the effective action of the lattice phonon mode
$ {\cal L}_L[\theta ] =( \partial_{\tau} \theta )^2 + a_L ( \partial_z \theta )^2  +  b_L  ( \partial_{\perp} \theta )^4 $.
At any finite temperatures, it becomes a algebraic ordered spin nematic state.

\begin{figure}[!t]
\begin{center}
\includegraphics[width=3.4in]{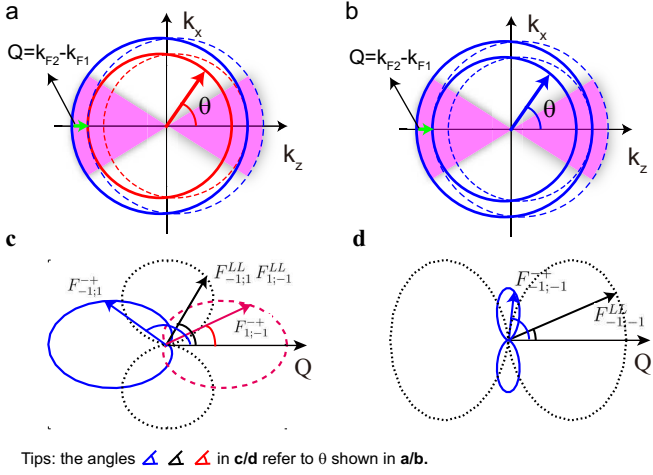}
\end{center}
\caption{\textbf{Fermi surface nesting and overlap factor.}
(\textbf{a}) The Fermi surfaces shown by the solid circles for small SOC ($\gamma\ll 1$), where the two Fermi surfaces have different helicities (different colors). The dashed circles are the ones after translation with the nesting momentum $Q=k_{F2}-k_{F1}$ with $k_{F1,2}$ the Fermi momenta. The angle $\theta$ shows the position at the Fermi surface. The Fermi surfaces inside the shaded regime around $\theta=0,\pi$ are well nested, i.e., the translated Fermi surfaces nearly overlap with the original ones.
(\textbf{b}) The Fermi surfaces shown by the solid circles for large SOC ($\gamma \gg 1$), where the two Fermi surfaces have the same helicities (blue colors). The conventions are the same as above.
(\textbf{c}) The overlap factor as a function of $\theta$ (denoting the position on the Fermi surface shown in \textbf{a}) in the transverse ($F^{-+}_{s;-s}$ with $s=\pm 1$)and longitudinal ($F^{LL}_{s;-s}$ with $s=\pm 1$) channels for small SOC in (\textbf{a}), where the the length of the three arrow lines ending at the red dashed line, black dotted line and the blue solid line correspond to the value of overlap factors as labeled nearby. The angles between the three arrows and $Q$ correspond to the position on the Fermi surface shown by \textbf{a}.
(\textbf{d}) The overlap factor in the transverse ($F^{-+}_{-1;-1}$) and longitudinal ($F^{LL}_{-1;-1}$) channels for large SOC in (\textbf{b}). All the conventions are the same as in (\textbf{c}).}
\label{fig:FS}
\end{figure}

In the $ \gamma \rightarrow \infty $ limit or equivalently zero density limit,
the system may become a Wigner crystal. In this limit, our RPA approach which in principle works for high density limit only breaks down.
A different approach, such as a scaling argument treating the interaction non-perturbatively, may be needed to work out
the true ground state in such a zero density limit.

\begin{figure}[!t]
\begin{center}
\includegraphics[width=3.4in]{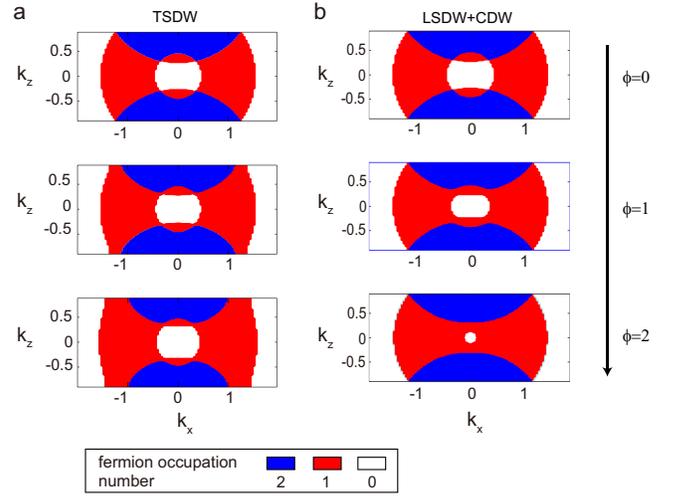}
\end{center}
\caption{\textbf{Fermi surfaces reconstruction.}
(\textbf{a}) The Fermi surface for SOC strength $\gamma=0.8238$ and orbital wave vector $Q=1.8 $ inside the TSDW phase.
From up to down, the order parameter is increasing as $\phi=0,1,2$.
The fermion occupation number in the purple, yellow, white regime are 2,1,0 respectively. We find
the intersection of Fermi surface at or inside the Brillouin zone boundary split due to the presence of the periodic TSDW order.
The interactions at $\vec{k}=0$ are preserved by symmetry, which is not shown in this case.
(\textbf{b}) As a comparison, we show the Fermi surface inside LSDW+CDW phase, where the values of SOC strength $\gamma$, orbital wave vector $Q$, the order parameters $\phi$ and the other conventions are the same as above. By contrast to TSDW phase, the intersection of Fermi surface at the Brillouin zone boundary is preserved by symmetry.}
\label{fig:FS}
\end{figure}

\textbf{Mechanism of the spin flop transition}
As we have shown above, the TSDW order turns into the LSDW order as the SOC strength $\gamma>1.5$, which is a spin flop transition. This change of the favorite SDW patterns for different SOC strengths can be well understood by the different helicity of Fermi surfaces in two limiting cases with small and large SOC, where the helicity is defined as the eigenvalue of the SOC term $\vec{s}\cdot \vec{k}/|\vec{k}|$. In both cases, the two Fermi surfaces are well nested with the nesting wavevector $\vec{Q}=(k_{F2}-k_{F1})\hat{e}_z$ (actually the direction of wavevector $\vec{Q}$ is arbitrary). And this nesting condition is best satisfied near $\theta=0,\pi$ where $\theta$ is the angle between the Fermi wavevector $\vec{k}_F$ and $\vec{Q}$ (schematically shown in Fig. 3 \textbf{a} and \textbf{b}). With SOC, the density and spin susceptibility is anisotropic due to the overlap factor $F^{\mu \nu}_{sr}$ defined in Eq. (\ref{densityspin}). For small SOC with $\gamma \rightarrow 0$, the two Fermi surfaces have different helicities ($\pm 1$). We find the overlap factor for the transverse susceptibility is $F^{-+}_{1;-1}=\cos^4 (\theta/2)$, $F^{-+}_{-1;1}=\sin^4 (\theta/2)$ which takes its maximum at $\theta=0,\pi$, i.e., the place where the nesting condition is best satisfied; the overlap factor for the longitudinal susceptibility is $F^{LL}_{-1;1}=F^{LL}_{1;-1}=\sin^2 \theta$ which is suppressed at $\theta=0,\pi$ (see Fig. 3\textbf{c}). As a result, the susceptibility is significantly enhanced in the transverse channel at the wavevector $|\vec{Q}|=k_{F2}-k_{F1}$ but suppressed in the longitudinal channel. For large SOC with $\gamma \rightarrow \infty$, the inner Fermi surface changes its helicity from $1$ to $-1$, the overlap factor for transverse susceptibility is $F^{-+}_{-1;-1}={1\over 4}\sin^2\theta$ and that for the longitudinal susceptibility is $F^{LL}_{-1;-1}=\cos^2 \theta$ (see Fig. 3 \textbf{d}). In this case, the susceptibility is significantly enhanced in the longitudinal channel at the wave vector $|\vec{Q}|=k_{F2}-k_{F1}$ but suppressed in the transverse channel. This is consistent with the numerical calculations of the static susceptibility for different SOC strengths (see Fig. 5 and Methods): the transverse susceptibility $\chi^{-+}$ reaches the maximum at $|\vec{Q}|=k_{F2}-k_{F1}$ for small SOC while the longitudinal susceptibility $\chi^{LL}$ reaches the maximum at $|\vec{Q}|=k_{F2}-k_{F1}$ for large SOC. So, in between there is a transition from TSDW to LSDW shown in the phase diagram. As shown in Fig. 2, the spin flop transition happens at $ \gamma \sim 1.5 $ and it is first order because $ Q_T > Q_L$.

\textbf{Fermi surface re-constructions}
The T/L-SDW order provides a periodic potential to fermions, which leads to enfoldment of Fermi surfaces according to the reciprocal lattice vector $ Q_{T/L}\hat{z} $ and subsequent Fermi surface reconstruction as shown in Fig. 4.
In the presence of SOC, there are two Fermi surfaces, which gives rise to two kinds of Fermi surface intersections: the one is that between the same Fermi surface which happens at the Brillouin zone boundary and that at $k_{z}=0$; the other
is that between the different Fermi surfaces, which usually happens inside the Brillouin zone but $k_{z}\neq 0$
(exceptions could happen but are not general cases).
We find the TSDW and LSDW order have quite different influence on the Fermi surface reconstructions: i) the Fermi surface intersection
at the Brillouin zone boundary will be opened in the presence of TSDW
order but keep in the presence of LSDW order; ii) the Fermi surface
intersection at the $\vec{k}=0$ axis will be kept in the presence of both
kinds of SDW orders; as a result, the Dirac cone structure is kept; iii) the Fermi surface interaction is opened
in general for other cases. The numerical calculation shown in Fig. 4 clearly displays different behaviors of Fermi surface reconstruction at the Brillouin zone boundary. These properties revealed here on the Fermi surface structure would be potentially valuable in further studies of itinerant fermions with the presence of T/L-SDW order.

The different Fermi surface reconstructions for T/L-SDW order can be illustrated by symmetry argument in an elegant way,
whose procedure is provided in the Methods. One can also perform simple degenerate perturbation calculations to find out
the band gap openings at the spectrum intersections. For instance, we find the gap opening
at the Brillouin zone boundary due to TSDW order is given by $\Delta^{T}_{BZ} = \phi_0 \sin^2 \frac{\theta}{2}$ where
$\theta<\pi/2$ is the angle between the Fermi momentum $\vec{k}_F$ ending at the Brillouin zone boundary and the reciprocal lattice vector $\vec{Q}_T$.
In sharp contrast to the TSDW phase, due to the different symmetry breaking patterns in the LSDW+CDW phase,
there is no gap opening at the Brillouin zone boundary $ \Delta^{L}_{BZ}=0 $.
There is another gap opening at the crossing between outer Fermi surface and inner Fermi surface
inside the Brillouin zone. First order degenerate perturbation shows that
$ \Delta^{T}_{I}= \phi_0 |\cos \frac{\theta_1}{2} \cos \frac{\theta_2}{2}| $ for TSDW order and $ \Delta^{L}_{I}= \phi_0 |\sin \frac{ (\theta_1-\theta_2)}{2}  | $ for LSDW order, where the $ \theta_{1,2} $ are the two polar angles of the crossing Fermi momenta with $ \vec{k}_{F2}-\vec{k}_{F1}= Q_T \hat{z}$ or $Q_L \hat{z}$.

\textbf{\large{Discussion}}
 
 In this Article, we have demonstrated that the repulsively interacting Fermi gas with a Weyl SOC provides a completely new system displaying a herd of novel properties, some of which are found general in spin-orbit coupled systems.
 Its Hamiltonian has a new class of spin-orbital coupled symmetries which is the decisive factor to give rise to innumerable important physical consequences.
 We establish exact relations on various density-spin correlation functions by exact symmetry analysis and
 point out that there is no FM phase in the Weyl SOC system.
 By using detailed calculations at the random phase approximation level, we found that the paramagnetic state will become
 unstable against itinerant stripe TSDW and LSDW with coexisting CDW and both of them could emerge at extremely
 weak interactions for strong SOC, which creates a favorable condition to observe these phases in experiments.
 The nature of these phase transitions is revealed by studying the quantum fluctuation effect near the quantum critical points.
 The itinerant TSDW and LSDW + CDW may display the anomalous Hall effect \cite{ahe} and will be investigated in a future publication.
 The results achieved here can be straightforwardly extended to 2D Rashba SOC case and have deep and wide implications
 in relevant condensed matter systems.

 According to present cold atom experiment, the repulsively interacting Fermi gas can be prepared by loading fermionic atoms such as $^{6}$Li atoms in Ref. \cite{GBJo} to the upper branch of a Feshbach resonance. This upper branch is an excited branch, therefore metastable due to near-resonant three-body recombination effect \cite{three1,three2,three3}. Fortunately, the itinerant magnetism in Fermi gas with a strong SOC could emerge at a much weaker interaction strength. Thus, the three-body recombination effect is so weak that magnetic domains could be hopefully observed in a time scale shorter than the lifetime of the upper branch. The preparation of Weyl SOC has been proposed by two schemes, i.e. light-matter coupling \cite{3DSOC1} and magnetic gradient methods \cite{3DSOC2,3DSOC3}. The latter scheme by reason of not facing the laser heating problem is a promising candidate. In this scheme, taking $^{6}$Li system as an example, the two hyperfine sublevels $|{1/2,-1/2} \rangle$ and $|{1/2,1/2} \rangle$ can be chosen as two pseudo-spin $1/2$ states.
Using $N\sim 10^4$ $^{6}$Li atoms inside an isotropic trap with a trap frequency $2\pi \times 10$Hz, and a typical magnetic field gradient strength $\nabla B=0.09G/\mu$m (within practical range \cite{Trinker}),  we estimate $\gamma
\sim 1.82$. Based on this, this system with the above parameters falls into the LSDW regime with the orbital momentum $Q_L\sim 0.13k_R$.
 The critical interaction strength $k_F a_c\sim 0.12$ has the order of magnitude smaller than the
 value $ \pi $ (Fig.1 \textbf{c} at $ \gamma=0 $) for a possible FM at the same density without SOC \cite{fmcold11,fmcold14,fmcold15,fmcold16}.
 Hence experiments could easily reach the very weak interaction regime to obtain both TSDW and LSDW+CDW phases.
 Once the Weyl SOC is implemented in cold atom experiments, all these novel phenomena can be detected by various
 established experimental techniques \cite{speckle1,speckle2,Shin,Gemelke}. For instance, one can detect the paramagnetic to both transverse and longitudinal SDW phase transitions by monitoring density profiles of each
 spin state using a double-shot phase contrast imaging technique \cite{Shin,Gemelke,double2} or measuring the density current,
 a signature of the T/L-SDW phase in spin-orbit coupled systems, by absorption-imaging the atom cloud after time of flight.
 One can also detect the dispersions of collective excitations with different interaction or SOC strength by Bragg
 spectroscopy \cite{collective1,bragg1,bragg3,bragg6,light1,light2}.

 \bigskip
 
 \textbf{\large{Methods}}
 
 \textbf{Path-integral formulism}
The interaction between fermions can be divided into the density and spin channel
$\mathcal{H}_I=\frac{g}{8}\int d\vec{r}[\rho(\vec{r})^2-\vec{S}(\vec{r})^2]$, where $\rho(\vec{r})=\Psi^{\dagger}\Psi(\vec{r})$ and $\vec{S}(\vec{r})= \Psi^{\dagger}\vec{\sigma}\Psi(\vec{r})$. With the presence of SOC, the density fluctuation is coupled with the spin fluctuation\cite{SSZhang}, therefore, we introduce the density-spin order parameter $\phi_{\mu},\mu=n,x,y,z$ to decouple the interaction term via a Hubbard-Stratonovich transformation. Integrating out the fermionic fields leads to the effective action for $\phi_{\mu}$:
\begin{eqnarray}\label{eq:SH}
\mathcal{S}=\int d^3\vec{r}\int_0^{1/T} d\tau \frac{1}{2g}\phi_{\mu}^2 -\text{Tr} \ln
\left( -\mathcal{G}^{-1}_0+ M\right),
\label{action}
\end{eqnarray}
where $\mathcal{G}^{-1}_0=-\partial_\tau-\mathcal{H}_{0}+\mu$, $M={i\over 2}\phi_{n}\sigma^{0}+{1\over 2} \vec{\phi}\cdot \vec{\sigma}$.

The density channel $\phi_n$ has a non-zero  imaginary saddle point value due to the finite particle density of the fermions
 which could be eliminated by re-defining $ \delta \phi_n$ as the deviation from its saddle point value. To the second order of $\delta \phi_{\mu}$,
 we obtain: $\mathcal{S}_H^{(2)}=1/(2\beta V)\sum_{\vec{q},n}\mathcal{K}^{\mu\nu} \delta \phi_{\mu}^*\delta \phi_{\nu}(\vec{q},i\omega_n)$ where $\mathcal{K}^{\mu\nu}=\delta^{\mu\nu}/g -\bar{\chi}^{\mu\nu}(\vec{q},i\omega_n)/4$, $\bar{\chi}^{\mu\nu}$ is related to the usual density-spin susceptibility $\chi^{\mu\nu}(\vec{q},i\omega_n)=\langle s^{\mu}(\vec{q},i\omega_n)s^{\nu}(-\vec{q},-i\omega_n) \rangle$ via $\bar{\chi}^{00}=-\chi^{00}$, $\bar{\chi}^{0i}=i\chi^{0i}$, $\bar{\chi}^{ij}=\chi^{ij}$ with $i,j,=x,y,z$.

 Adding density-spin sources $h_{\mu}$ to Eq. (\ref{eq:SH}) and integrating out $ \delta \phi_{\mu} $,
one can obtain the dynamic density-spin response functions at the random phase approximation level through $\chi_{RPA}^{\mu\nu}=\delta^2\mathcal{S}_H/\delta h_{\mu}\delta h_{\nu}$. The concrete expression has the conventional form $\chi_{RPA}=\chi_0(1+g/4\eta \chi_0)^{-1}$ where $\eta=diag\{1,-1,-1,-1\}$ (i.e., the Minkowski matric) and the non-interacting density-spin susceptibilities are given by
\begin{equation}
\chi^{\mu \nu }_{0}(\vec{q},i\omega_n)=\frac{1}{V}\sum_{k,sr} F_{sr}^{\mu \nu }(k+q,k)\frac{
n_{F}(\xi _{k+q,s})-n_{F}\left( \xi _{k,r}\right) }{i\omega_n -\left( \xi
_{k+q,s}-\xi _{k,r}\right)},
\label{densityspin}
\end{equation}
where $\xi _{k,s}$ is the fermion spectrum with $s$ the helicity $\hat{p}\cdot \vec{\sigma}|ps\rangle=s|ps\rangle$,
$n_{F}\left( \xi _{k,s}\right) $ the Fermi distribution function, $\xi _{k+q,s}-\xi _{k,r}$ the particle-hole excitation energy and $F_{sr}^{\mu \nu }(p,q)=\langle ps| \sigma^{\mu}|qr\rangle\langle qr|\sigma^{\nu}|ps\rangle$
the overlap factor. The poles of $\chi_{RPA}$ determined by $\det(1+g/4\eta \chi_0)=0$ give rise to the collective excitations above the PM state. Taking the static limit, $i\omega_n \rightarrow 0$, one obtains the static density and spin susceptibilities. From the symmetry analysis, we deduce that the static susceptibility can be decoupled to three independent channels: the density channel $\chi^{nn}(q)$, the longitudinal spin channel $\chi^{LL}(q)$ and the transverse spin channels $\chi^{+-}(q)$ and $\chi^{-+}(q)$. Note that the density channel is decoupled with the longitudinal spin channel in the static limit. The divergences of the static susceptibilities indicate instabilities of the system in corresponding channels mentioned above, and their competitions have been discussed in the main body of text.

The non-interacting density-spin susceptibilities $\chi^{\mu \nu }_{0}$ can be carried out analytically in two cases: i) the singular form near $(\omega,\vec{q})=(0,\vec{0})$ where $\chi^{\mu \nu }_{0}$ is the function of the ratio $y=\omega/v_F q$ with $v_F$ the Fermi velocity;
ii) the uniform susceptibility $\chi^{\mu\nu}(\omega,\vec{0})$. These results could provide RPA solutions to the velocity of the gapless sound mode and the energy gaps of the gapped modes. Please see Supplementary Material for more details.

\begin{figure}[!t]
\begin{center}
\includegraphics[width=3.4in]{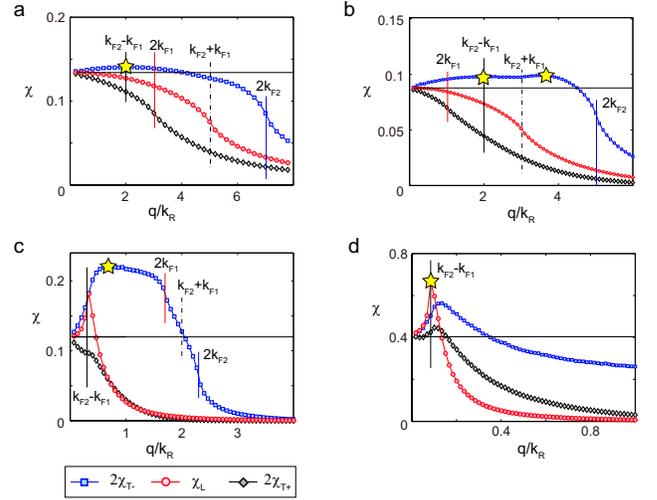}
\end{center}
\caption{ \textbf{Spin susceptibility.}
(\textbf{a})-(\textbf{d}) The spin susceptibility for the longitudinal spin (red line with circles) and two spiral transverse spins (blue line with squares: $\chi_{T-}$, black line with diamonds: $\chi_{T+}$) of three dimensional non-interacting Fermi gas with a Weyl SOC. The SOC strength is taken as $\gamma=0.35,0.5,1.3,2$, respectively. The black horizonal line is the analytical value of the static and uniform susceptibility with $q=0$. The various vertical lines represent different momenta $2k_{F1},2k_{F2},k_{F2}+k_{F1},k_{F2}-k_{F1}$ formed by the Fermi momenta. Within random phase approximation, the system becomes unstable if the maximum of the non-interacting susceptibilities in corresponding channels equal to $4/g$ with $g$ the repulsive s-wave interaction strength. We can find that the system would be unstable in the transverse spin channel first for \textbf{a}-\textbf{c} and turns to the longitudinal channel in \textbf{d}. We have marked the maximum of the susceptibility by yellow stars. We find in \textbf{b} that there are two maximum and they compete with each other with increasing SOC, which gives rise to the jump of the orbital wavevector $Q$ as shown in Fig. 2b. As analyzed in the main text, the susceptibility in the transverse (longitudinal) channel takes its maximum at $q=k_{F2}-k_{F1}$ for small (large) SOC strength.}
\label{fig:FS}
\end{figure}

\textbf{Random phase approximation and static susceptibility}
The symmetry analysis in the main body of text indicates that there are three competing instabilities given by: $\chi_{+-}^{-1}=0$, $\chi_{-+}^{-1}=0$ and $\chi_L^{-1}=0$. Within random phase approximation (RPA) which corresponds to the Gaussian fluctuation level in the path-integral formulism developed above, the susceptibilities are given by $\chi_{+-}^{RPA}=\chi_{+-}^0/(1-g/2\chi_{+-}^0)$, $\chi_{-+}^{RPA}=\chi_{-+}^0/(1-g/2\chi_{-+}^0)$ and $\chi_{L}^{RPA}=\chi_{L}^0/(1-g/4\chi_{L}^0)$. Instabilities of the paramagnetic state would happen when $2\chi_{+-}^0$, $2\chi_{-+}^0$ or $\chi_L^0$ equal to $4/g$. We find no clean analytical expressions for the static susceptibilities similar to the Lindhard function for systems without SOC. In Fig. 5, we show the numerical calculation of the susceptibility of the non-interacting Fermi gas with a Weyl SOC for different SOC strengths. We can clearly find the competition between different channels with increasing SOC which gives rise to the spin flop transition from transverse to longitudinal SDW phase. We can also find the transverse susceptibility $\chi^{-+}$ has a maximum at $q=k_{F2}-k_{F1}$ in \textbf{a} and then another maximum between $k_{F2}+k_{F1}$ and $2k_{F2}$ develops shown in \textbf{b}. Finally the newly emerged maximum exceeds the original one and gives rise to a jump of the orbital wavevector $Q$ discussed in the main text.

\textbf{Symmetry argument on the energy band structure}
The Fermi surface reconstruction or the energy band structure discussed in the main body can be illustrated by symmetry argument.
Let's first summarize the remaining symmetry of two types of SDW orders. We find that
both the T-SDW and L-SDW order are invariant up to a different corresponding
translation $T_{r}\left( a\right) $ with $a$ the distance along $\hat{z}$
direction under the following symmetry operations: 1) $\left[ SO(2)_{z}\times U\left(
1\right) _{z}\right] _{D}$; 2) $\mathcal{P}_{x},\mathcal{P}_{y}$, $\mathcal{P%
}_{z}$; 3) the time reversal symmetry. Besides, the translational symmetry
is broken to a discrete symmetry which leads to the band structure.

For T-SDW order assumed to be $\vec{\phi}=\phi \left( \cos Qz,-\sin
Qz,0\right) $, continuous rotational symmetry about $z$ axis leads to that
the spectrum is rotational symmetric about the $k_{z}$ axis. Therefore, we
need only to consider the case with $k_{y}=0$. The T-SDW order preserves
the $\mathcal{P}_{x}$ and the $\mathcal{P}_{y}^{^{\prime }}=\mathcal{P}
_{y}+T_{r}(\frac{\pi }{Q})$ symmetry where the prime in the subscript
indicates an combined translation.
The $\mathcal{P}_{x}$ ($\mathcal{P}_{y}^{^{\prime }}$) symmetry acting on
the single particle eigenstates with given momentum $\vec{k}$ in the first Brillouin zone
will be closed only when $k_{z}=-Q/2$ (Brillouin zone boundary) or $k_{z}=0$.
In concrete, we assume the eigen-wave function at the Brillouin zone boundary to be:
\begin{equation}
|\Psi \rangle _{BZ}=\sum_{n;\sigma }a_{n\sigma }|\frac{2n+1}{2}Q,\sigma \rangle,  \label{eq:WFBZ}
\end{equation}
where ${2n+1\over 2}Q$ represents the $k_z$ component of $\vec{k}$, $\sigma=1,2$ the index of spin component. Here and after, we ignore the $k_{x,y}$ component in our representation of single particle state $|\vec{k},\sigma\rangle$ unless it is specialized. By the same convention, we assume the wavefunction at $k_z=0$ to be:
\begin{equation}
|\Psi \rangle _{0}=\sum_{n;\sigma }b_{n\sigma }|nQ,\sigma \rangle. \label{eq:WF0}
\end{equation}
Since the Hamiltonian is real, the coefficients $a_{n\sigma }$ and $b_{n\sigma}$ can be chosen to be real values.
Now, applying the $\mathcal{P}_{x}$ symmetry to the wave functions $|\Psi\rangle_{BZ}$ and $|\Psi\rangle _0$ twice leads to: $
\mathcal{P}_{x}^{2}|\Psi \rangle_{BZ}=|\Psi \rangle _{BZ}$ and $\mathcal{P}_{x}^{2}|\Psi \rangle _{0}=|\Psi \rangle _{0}$. This property is
not sufficient to guarantee the two-fold degeneracy at both the Brillouin zone boundary and $\vec{k}=0$.
Then we consider the $\mathcal{P}_{y}^{^{\prime }}$ symmetry, which needs to further take $k_x=0$ (namely the place at the center of the Brillouin zone boundary and $\vec{k}=\vec{0}$), and carry out the same procedure as above. We find and $\mathcal{P}_{y}^{^{\prime }2}|\Psi \rangle _{BZ}=|\Psi \rangle _{BZ}$ and $\mathcal{P}_{y}^{^{\prime }2}|\Psi\rangle _{0}=-|\Psi \rangle _{0}$. The later relation is non-trivial to lead to the two-fold degeneracy of the energy spectrum at $\vec{k}=0$. As a result, the Fermi surface interaction at $\vec{k}=0$ and Dirac cone structure are preserved after the TSDW order turns on.

The band structure in the presence of LSDW order can be considered in the same way.
The order parameter assumed as $\vec{\phi}=\phi \left( 0,0,\cos Qz\right) $ is invariant under both the $
\mathcal{P}_{x}^{^{\prime }}=\mathcal{P}_{x}+T_{r}(\frac{\pi }{Q})$ and $%
\mathcal{P}_{y}^{^{\prime }}=\mathcal{P}_{y}+T_{r}(\frac{\pi }{Q})$
symmetry. The eigen-wave functions at the Brillouin zone boundary and $\vec{k}=0$ are given by Eqs. (\ref{eq:WFBZ}) and (\ref{eq:WF0}), respectively.
Since the Hamilntonian is real, all of coefficients $a_{n\sigma}$ and $b_{n\sigma}$ can be chosen to be real. Applying the $\mathcal{P}_{x}^{^{\prime }}$
symmetry, we find $\mathcal{P}_{x}^{^{\prime }2}|\Psi \rangle _{BZ}=-|\Psi \rangle _{BZ}$ and $\mathcal{P}_{x}^{^{\prime }2}|\Psi \rangle _{0}=|\Psi \rangle _{0}$. The former relation is non-trivial and leads to the two-fold degeneracy of the energy
spectrum at the Brillouin zone boundary. We consider again the $\mathcal{P}_{y}^{^{\prime }}$ symmetry, which is the same as what we have encoutered
in the case of TSDW order: it guarantees the two-fold degeneracy at $\vec{k}=0$ but not sufficient at $\left( k_{x},k_{z}\right) =\left( 0,-Q/2\right) $.
The Dirac cone structure is also preserved in the presence of LSDW order.

As we have shown in the main body, the LSDW order would induce a CDW order with
orbital momentum $2Q$. Because the density order is invariant under the
above symmetry operations, all of these results would not change. Any other
places of the Brillouin zone would not be protected by symmetry, therefore,
there would be band gap opening there when the T/L-SDW order is engineered.

\textbf{Acknowledgments}
This work was supported by the NKBRSFC under grants Nos. 2011CB921502, 2012CB821305, NSFC under grants Nos. 61227902, 61378017, 11434015, and SPRPCAS under grants No. XDB01020300. J. Ye was supported by NSF-DMR-1161497, NSFC-11174210.


\begin{thebibliography}{99}

\bibitem{stone}
Stoner, E.
Collective electron ferromagnetism.
\textit{Philos. Mag.} \textbf{15}, 1018 (1933).



\bibitem{millis1}
Millis, A. J.
Effect of a nonzero temperature on quantum critical points in itinerant fermion systems.
\textit{Phys. Rev. B} \textbf{48}, 7183 (1993).

\bibitem{millis2}
Z\"{u}licke, U. and Millis, A. J.
Specific heat of a three-dimensional metal near a zero-temperature magnetic phase transition with dynamic exponent z=2, 3, or 4.
\textit{Phys. Rev. B} \textbf{51}, 8996 (1995).

\bibitem{stoner1}
Sogo, T. and Yabu, H.
Collective ferromagnetism in two-component Fermi-degenerate gas trapped in a finite potential.
\textit{Phys. Rev. A} \textbf{66}, 043611 (2002).

\bibitem{stoner2} Duine, R. A. and MacDonald, A. H.
Itinerant ferromagnetism in an ultracold atom fermi gas.
\textit{Phys. Rev. Lett.} \textbf{95}, 230403 (2005).

\bibitem{stoner3}
LeBlanc, L. J., Thywissen, J. H., Burkov, A. A. and Paramekanti, A.
Repulsive Fermi gas in a harmonic trap: Ferromagnetism and spin textures.
\textit{Phys. Rev. A} \textbf{80}, 013607 (2009).

\bibitem{stoner4}
Dong, H., Hu, H., Liu, X. -J. and Drummond, P. D.
Mean-field study of itinerant ferromagnetism in trapped ultracold Fermi gases: Beyond the local-density approximation.
\textit{Phys. Rev. A} \textbf{82}, 013627 (2010).


\bibitem{corre1}
Gutzwiller, M. C.
Effect of correlation on the ferromagnetism of transition metals.
\textit{Phys. Rev. Lett.} \textbf{10}, 159 (1963).


\bibitem{corre3}
Vilk, Y. M., Chen, L. and Tremblay, A. -M. S.
Theory of spin and charge fluctuations in the Hubbard model.
\textit{Phys. Rev. B} \textbf{49}, 13267 (1994).

\bibitem{corre4} Zhai, H.
Correlated versus ferromagnetic state in repulsively interacting two-component Fermi gases.
\textit{Phys. Rev. A} \textbf{80}, 051605(R) (2009).

\bibitem{corre5}
He, L. and Huang, X. -G.
Nonperturbative effects on the ferromagnetic transition in repulsive Fermi gases
\textit{Phys. Rev. A} \textbf{85}, 043624 (2012).


\bibitem{corre7} Auerbach, A.
\textit{Interacting electrons and quantum magnetism}, Springer-Verlag (1998).


\bibitem{GBJo} Jo, G. B., {\sl et al.}
Itinerant ferromagnetism in a Fermi gas of ultracold atoms.
\textit{Science} \textbf{325}, 1521 (2009).


\bibitem{fmcold11}
Duine, R. A. and MacDonald, A. H.
Itinerant ferromagnetism in an ultracold atom Fermi gas.
\textit{Phys. Rev. Lett.} \textbf{95}, 230403 (2005).

\bibitem{fmcold14}
Pilati, S., Bertaina, G., Giorgini, S. and Troyer, M.
Itinerant ferromagnetism of a repulsive atomic Fermi gas: A quantum Monte Carlo study.
\textit{Phys. Rev. Lett.} \textbf{105}, 030405 (2010).

\bibitem{fmcold15}
Cui, X. and Ho, T.-L.
Phase separation in mixtures of repulsive Fermi gases driven by mass difference.
\textit{Phys. Rev. Lett.} \textbf{110}, 165302 (2013).

\bibitem{fmcold16}
Cui, X. and Ho, T.-L.
Ground-state ferromagnetic transition in strongly repulsive one-dimensional Fermi gases.
\textit{Phys. Rev. A} \textbf{89}, 023611 (2014).


\bibitem{two1}
He, L.
Finite range and upper branch effects on itinerant ferromagnetism in repulsive Fermi gases: Bethe¨CGoldstone ladder resummation approach.
\textit{Annals of Physics}, \textbf{351}, 477-503 (2014).

\bibitem{two2}
He, L., Liu, X. -J., Huang, X. -G. and Hu, H.
Stoner ferromagnetism of a strongly interacting Fermi gas in the quasirepulsive regime.
\textit{arXiv:} 1412.2412.



\bibitem{polaron2}
Massignan, P., Yu, Z. and Bruun, G. M.
Itinerant ferromagnetism in a polarized two-component Fermi gas.
\textit{Phys. Rev. Lett.} \textbf{110}, 230401 (2013).

\bibitem{polaron3}
Massignan, P., Zaccanti, M. and Bruun, G. M.
Polarons, dressed molecules and itinerant ferromagnetism in ultracold Fermi gases.
\textit{Rep. Prog. Phys.} \textbf{77} 034401 (2014).

\bibitem{pband1}
Zhang, S., Huang, H. and Wu, C.
Proposed realization of itinerant ferromagnetism in optical lattices.
\textit{Phys. Rev. A} \textbf{82}, 053618 (2010).





\bibitem{revSOC1}
Dalibard, J., Gerbier, F., Juzeli\=unas, G. and \"{O}hberg, P.
\emph{Colloquium:} Artificial gauge potentials for neutral atoms.
\textit{Rev. Mod. Phys.} \textbf{83}, 1523 (2011).

\bibitem{revSOC2}
Galitski, V. and Spielman, I. B.
Spin-orbit coupling in quantum gases.
Nature \textbf{494}, 49-54 (2013).


\bibitem{revSOC4}
Zhai, H.
Degenerate quantum gases with spin¨Corbit coupling: a review.
\textit{Rep. Prog. Phys.} \textbf{78}, 026001 (2015).



\bibitem{SOCexp1}
Lin, Y. J., Jim\'{e}nez-Garc\'{i}a, K. and Spielman, I. B.
Spin-orbit-coupled Bose-Einstein condensates.
Nature (London) \textbf{471}, 83 (2011).

\bibitem{SOCexp2}
Wang, P., {\sl et al.}
Spin-orbit coupled degenerate Fermi gases.
\textit{Phys. Rev. Lett.} \textbf{109}, 095301 (2012).

\bibitem{SOCexp3}
Cheuk, L. W., {\sl et al.}
Spin-injection spectroscopy of a spin-orbit coupled Fermi gas.
\textit{Phys. Rev. Lett.} \textbf{109}, 095302 (2012).




\bibitem{soctheory31}
Yu, Z. Q. and Zhai, H.
Spin-orbit coupled Fermi gases across a Feshbach resonance.
\textit{Phys. Rev. Lett.} \textbf{107}, 195305 (2011).

\bibitem{soctheory32}
Hu, H., Jiang, L., Liu, X. -J. and Pu, H.
Probing anisotropic superfluidity in atomic Fermi gases with Rashba spin-orbit coupling.
\textit{Phys. Rev. Lett.} \textbf{107}, 195304 (2011).

\bibitem{soctheory33}
Qu, C., {\sl et al.}
Topological superfluids with finite-momentum pairing and Majorana fermions.
\textit{Nat. Commun.} \textbf{4}, 2710 (2013).

\bibitem{soctheory34}
Zhang, W. and Yi, W.
Topological Fulde-Ferrell-Larkin-Ovchinnikov states in spin-orbit-coupled Fermi gases.
\textit{Nat. Commun.} \textbf{4}, 2711 (2013).

\bibitem{soctheory35}
Vyasanakere, J. P. and Shenoy, V. B.
Bound states of two spin-1/2 fermions in a synthetic non-Abelian gauge field.
\textit{Phys. Rev. B} \textbf{83}, 094515 (2011).

\bibitem{soctheory36}
Vyasanakere, J. P., Zhang, S. and Shenoy, V. B.
BCS-BEC crossover induced by a synthetic non-Abelian gauge field.
\textit{Phys. Rev. B} \textbf{84}, 014512 (2011).

\bibitem{soctheory37}
Zhou, Qi and Cui, X.
Fate of a Bose-Einstein condensate in the presence of spin-orbit coupling.
\textit{Phys. Rev. Lett.} \textbf{110}, 140407.

\bibitem{soctheory38}
Dong, Y., Dong, L., Gong, M. and Pu, H.
Dynamical phases in quenched spin-orbit-coupled degenerate Fermi gas.
\textit{Nat. Commun.} \textbf{6}, 6103 (2015).


\bibitem{pairing} Yi-Xiang, Y., Ye, J. and Liu, W. -M.
Coherence lengths in attractively interacting Fermi gases with spin-orbit coupling.
\textit{Phys. Rev. A} \textbf{90}, 053603  (2014).

\bibitem{CJWu0}
Li, Y. and Wu, C.
Spin-orbit coupled Fermi liquid theory of ultracold magnetic dipolar fermions.
\textit{Phys. Rev. B} \textbf{85}, 205126 (2012).

\bibitem{SSZhang}
Zhang, S. S., Yu, X. L., Ye, J. and Liu, W. M.,
Collective modes of spin-orbit-coupled Fermi gases in the repulsive regime.
\textit{Phys. Rev. A} \textbf{87}, 063623 (2013).

\bibitem{topo}
Sun, F., {\sl et al.}
Topological quantum phase transition in synthetic non-abelian gauge potential: gauge invariance and experimental detections.
\textit{Scientific Reports} \textbf{3}, 2119 (2013).

\bibitem{rhboson}
Sun, F., Ye, J. and Liu, W. -M.
Rotated ferromagnetic Heisenberg model.
\textit{arXiv:}1408.3399.










\bibitem{bragg1}
Kozuma, M., {\sl et al.}
Coherent splitting of Bose-Einstein condensed atoms with optically induced Bragg diffraction.
\textit{Phys. Rev. Lett.} \textbf{82}, 871 (1999).


\bibitem{bragg3}
Stamper-Kurn, D. M., {\sl et al.}
Excitation of phonons in a Bose-Einstein condensate by light scattering.
\textit{Phys. Rev. Lett.} \textbf{83}, 2876 (1999).



\bibitem{collective1}
Ji, S-C., {\sl et al.}
Softening of roton and phonon modes in a Bose-Einstein condensate with spin-orbit coupling.
\textit{Phys. Rev. Lett.} \textbf{114}, 105301 (2015).

\bibitem{bragg6}
Ernst, P. T., {\sl et al.}
Probing superfluids in optical lattices by momentum-resolved Bragg spectroscopy.
\textit{Nat. Phys.} \textbf{6}, 56 (2010).

\bibitem{light1}
Ye, J., {\sl et al.}
Light-scattering detection of quantum phases of ultracold atoms in optical lattices.
\textit{Phys. Rev. A} \textbf{83}, 051604(R) (2011).

\bibitem{light2} Ye, J., {\sl et al.}
Optical Bragg, atomic Bragg and cavity QED detections of quantum phases and excitation spectra of ultracold atoms in bipartite and frustrated optical lattices.
\textit{Ann. Phys.} \textbf{328}, 103-138 (2013).

\bibitem{speckle1}
Sanner, C., {\sl et al.}
Speckle imaging of spin fluctuations in a strongly interacting Fermi gas.
\textit{Phys. Rev. Lett.} \textbf{106}, 010402 (2011).

\bibitem{speckle2} Sanner, C., {\sl et al.}
Correlations and pair formation in a repulsively interacting Fermi gas.
\textit{Phys. Rev. Lett.} \textbf{108}, 240404 (2012).


\bibitem{Shin}
Shin, Y., Schunck, C. H., Schirotzek, A. and Ketterle, W.
Observation of phase separation in a strongly interacting imbalanced Fermi gas.
\textit{Nature (London)} \textbf{451}, 689 (2008).

\bibitem{Gemelke} Gemelke, N., Zhang, X., Huang, C. -L. and Chin, C.
In situ observation of incompressible Mott-insulating domains in ultracold atomic gases.
\textit{Nature (London)} \textbf{460}, 995 (2009).





\bibitem{3DSOC1} Anderson, B. M., Juzeliunas, G., Galitski, V. M. and Spielman, I. B.
Synthetic 3D spin-orbit coupling.
\textit{Phys. Rev. Lett.} \textbf{108}, 235301 (2012).

\bibitem{3DSOC2} Anderson, B. M., Spielman, I. B. and Juzeli\=unas, G.
Magnetically generated spin-orbit coupling for ultracold atoms.
\textit{Phys. Rev. Lett.} \textbf{111}, 125301 (2013).

\bibitem{3DSOC3} Xu, Z. -F., You, L. and Ueda, M.
Atomic spin-orbit coupling synthesized with magnetic-field-gradient pulses.
\textit{Phys. Rev. A} \textbf{87}, 063634 (2013).




\bibitem{roton01}
Ye, J.,
Ginzburg-Landau theory of a supersolid.
\textit{Phys. Rev. Lett.} \textbf{97}, 125302 (2006).

\bibitem{roton02} Ye, J.,
Elementary excitation in a supersolid.
\textit{Europhysics Letters}, \textbf{82} 16001 (2008).

\bibitem{roton11}
Ye, J and Jiang, L.
Quantum phase transitions in bilayer quantum Hall systems at a total filling factor $\nu_T=1$.
\textit{Phys. Rev. Lett.} \textbf{98}, 236802 (2007).

\bibitem{roton12}
Ye, J.
Fractional charges and quantum phase transitions in imbalanced bilayer quantum Hall systems
\textit{Phys. Rev. Lett.} \textbf{97}, 236803 (2006).



\bibitem{LO1} Jiang, L. and Ye, J.
Lattice structures of Fulde-Ferrell-Larkin-Ovchinnikov state.
\textit{Phys. Rev. B} \textbf{76}, 184104 (2007).

\bibitem{LO2} Radzihovsky, L.
Fluctuations and phase transitions in Larkin-Ovchinnikov liquid-crystal states of a population-imbalanced resonant Fermi gas.
\textit{Phys. Rev. A} \textbf{84}, 023611 (2011).

\bibitem{book:Lubensky}
Chaikin, P. M. and Lubensky, T. C.
\textit{Principles of condensed matter physics} (Cambridge University Press, 2000).















\bibitem{ek1} Zachar, O., Kivelson. S. A. and Emery, V. J.
Landau theory of stripe phases in cuprates and nickelates.
\textit{Phys. Rev. B} \textbf{57}, 1422 (1998).

\bibitem{ek2} Kivelson, S. A., {\sl et al.}
How to detect fluctuating stripes in the high-temperature superconductors.
\textit{Rev. Mod. Phys.} \textit{75}, 1201 (2003).

\bibitem{ahe} Ye, J., {\sl et al.}
 Berry phase theory of the anomalous Hall effect: Application to colossal magnetoresistance manganites.
\textit{Phys. Rev. Lett.} \textbf{83}, 3737 (1999).

\bibitem{three1} Shenoy, V. B. and Ho, T. -L.
Nature and properties of a repulsive Fermi gas in the upper branch of the energy spectrum.
\textit{Phys. Rev. Lett.} \textbf{107}, 210401 (2011).

\bibitem{three2} Pekker, D., {\sl et al.}
Competition between pairing and ferromagnetic instabilities in ultracold Fermi gases near Feshbach resonances.
Phys. Rev. Lett. \textbf{106}, 050402 (2011).

\bibitem{three3} Weber, T., {\sl et al.}
 Three-body recombination at large scattering lengths in an ultracold atomic gas.
\textit{Phys. Rev. Lett.} \textbf{91}, 123201 (2003).

\bibitem{Trinker} Trinker, M., {\sl et al.}
Logarithmic periodicities in the bifurcations of type-I intermittent chao.
Appl. Phys. Lett. \textbf{92}, 254102 (2008).

\bibitem{double2} Lee, Y.-R., {\sl et al.}
Compressibility of an ultracold Fermi gas with repulsive interactions.
\textit{Phys. Rev. A} \textbf{85}, 063615 (2012).


\end{thebibliography}
\end{document}


\title{Supplementary Material for ``Itinerant magnetism in Weyl spin-orbit coupled Fermi gas''}
\author{Shang-Shun Zhang$^{1,2}$}
\author{Jinwu Ye$^{2,3}$}
\author{Wu-Ming Liu$^1$}
\affiliation{ $^{1}$Beijing National Laboratory for Condensed Matter Physics, Institute of
Physics, Chinese Academy of Sciences, Beijing 100190, China \\
$^{2}$ Department of Physics and Astronomy, Mississippi State University, MS 39762, USA  \\
$^{3}$ Key Laboratory of Terahertz Optoelectronics, Ministry of Education, Department of Physics, Capital Normal University, Beijing 100048, China
}
\date{\today}
\maketitle

In this Supplementary Material, we will first classify the symmetries of the Weyl SOC Hamiltonian Eq. (M1) of the main text and also
find their exact constraints on various density-spin correlation functions in Sec. I.
Then, in Sec.II, we make random phase approximation (RPA) calculations to find the quantitative behaviors of the collective modes through the poles of density-spin response functions in the paramagnetic side of this system.
Through out this Material, we use M to stand for the figures and equations in the main text.

\section{Symmetry analysis}
The symmetry of this system has been outlined in the first section of the main text, which leads to rigorous constraints on the density and spin susceptibilities as already shown there. Here, we provide some details on this symmetry analysis. Without loss of generality, we can pick up $\vec{q}$ as the $ \hat{z} $ axis, i.e., $\vec{q}= q_z \hat{z}$ with $q_z\in (-\infty,\infty)$. Then, the helical basis shown in Fig. M1a is just given by $\hat{T}_1=\hat{x},\hat{T}_2=\hat{y}$ and $\hat{L}=\hat{z}$. Note that we use a unified bases for $\pm \vec{q}$ here. The $[SU(2)_{spin} \times SO(3)_{orbit} ]_D $ symmetry along this $ \hat{z} $ axis dictates that
\begin{equation}
  R^{T}_{z} \chi^{\mu \nu} R_z= \chi^{\mu \nu}.
\end{equation}
 Expanding it leads to the fact that
 the dynamical $ 4 \times 4 $ density-spin response function $ \chi^{\mu\nu}(\vec{q}, \omega ) $ can be split into two $ 2 \times 2 $ subspaces defined in the main text:
 i) $T_1-T_2$ subspace with $ \chi^{xx}= \chi^{yy}, \chi^{xy}= -\chi^{yx} $; ii) $n$-$L$ subspace.

{\sl (1) In the $T_1$-$T_2$ subspace }

Under the $s^{\pm}=s^x\pm i s^y$ bases via a unitary transformation, it can be shown that in the  $T_1-T_2$ subspace:
\begin{equation}
 \chi=\begin{pmatrix}
		\chi^{+-} & 0 \\
		0 & \chi^{-+} \\
	\end{pmatrix},
\label{pm}
\end{equation}
where $ \chi^{+-}= 2 (\chi^{xx}- i  \chi^{xy} ), \chi^{-+}= 2 (\chi^{xx}+ i  \chi^{xy} ) $. The $ \mathcal{P}_x $ (or $\mathcal{P}_y $) symmetry dictates
\begin{equation}
\chi^{+-}(\vec{q}, \omega)=  \chi^{-+}(-\vec{q}, \omega).
\label{px}
\end{equation}
The $\mathcal{T}$ symmetry leads to
\begin{equation}
 \chi^{+-}(\vec{q}, \omega)=  \chi^{-+}(-\vec{q}, \omega).
\label{t}
\end{equation}
which is identical to Eq. (\ref{px}). Eq. (\ref{px}) leads to that $ \chi^{xx}=\chi^{yy} $ is an even function of $ q_z $,
 $ \chi^{xy}=-\chi^{yx} $ is an odd function of $ q_z $. Directly taking complex conjugate leads to:
\begin{equation}
[\chi^{+-}(\vec{q}, \omega)]^{*}=  \chi^{-+}(-\vec{q}, -\omega).
\label{cc}
\end{equation}
  which, combining with Eq. (\ref{px}), leads to
\begin{eqnarray}
 Re \chi^{+-}(\vec{q}, \omega)& = &   Re \chi^{+-}(\vec{q}, -\omega),   \nonumber   \\
 Im \chi^{+-}(\vec{q}, \omega) & = & - Im \chi^{+-}(\vec{q}, -\omega).
\label{ccri}
\end{eqnarray}

\begin{figure}[!h]
\begin{center}
\includegraphics[width=2.5in]{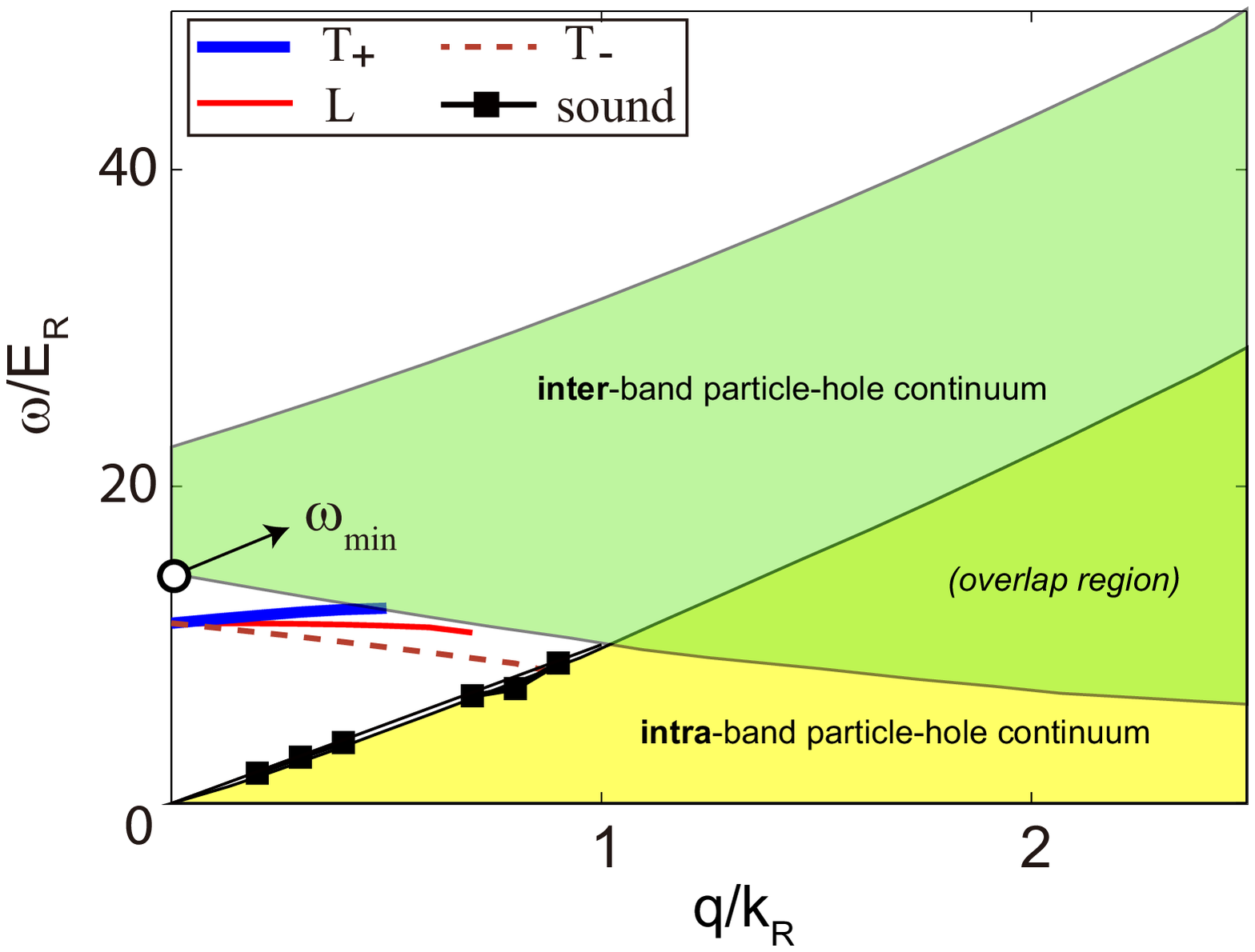}
\end{center}
\caption{(Color online)
Collective modes for 3D Fermi gas with Weyl-type SOC in the normal side.
The parameters using here are: $ \mu/E_R= 20 $ (equivalent to $ \gamma=0.28 $) which corresponds to the case $\gamma$) in Fig. 2
and $ g=0.9 g_c $ with $g_c$ the critical interaction strength of the putative ferromagnetic phase transition. The green (yellow) region is the intra-band  (inter-band)
particle-hole excitation continuum. $\omega_{min}$ denotes the lower edge of inter-band particle-hold continuum, which is taken as the unit in Fig. 2(d).}
\label{fig:coll2}
\end{figure}

Requiring $\chi^{\mu\nu}(\vec{q},\omega)$ to be Hermite leads to that $ \chi^{+-}, \chi^{-+} $ are real, therefore, Eq. (\ref{ccri}) gives rise to
the fact that $ \chi^{xx}=\chi^{yy} $ and $ \chi^{xy}=-\chi^{yx} $ are all even function of $ \omega $. The poles of $\chi^{+-}$ and $\chi^{-+}$ lead to the two gapped transverse modes $T_{\pm}$ respectively. In principle, $\chi^{+-}, \chi^{-+}$ might also develop imaginary parts, i.e., $Im \chi^{+-}(\vec{q}, \omega) \neq 0$, which represents the damping of the collective modes. At RPA level as presented in the next section, the damping is due to the coupling with the particle-hole (PH) excitations. The regime where $Im \chi^{+-}(\vec{q}, \omega) \neq 0$ is given by the intra- or inter- PH continuum as shown by the shaded regime in Fig. \ref{fig:coll2}.

{\sl (2) In the $ n-L $ subspace }

The $ \mathcal{P}_x $ (or $ \mathcal{P}_y $) symmetry dictates
\begin{eqnarray}
 \chi^{nn}(\vec{q}, \omega) & = &  \chi^{nn}(-\vec{q}, \omega), \chi^{LL}(\vec{q}, \omega)=  \chi^{LL}(-\vec{q}, \omega),
     \nonumber  \\
 \chi^{nL}(\vec{q}, \omega) & = & - \chi^{nL}(-\vec{q}, \omega),  \chi^{Ln}(\vec{q}, \omega)= - \chi^{Ln}(-\vec{q}, \omega),
\label{pxnl}
\end{eqnarray}
   which shows $ \chi^{nn},  \chi^{LL} $ are even function of $ q_z $, while
    $ \chi^{nL}, \chi^{nL}$ are an odd function of $ q_z $. The $\mathcal{T}$ symmetry leads to
\begin{eqnarray}
  \chi^{nn}(\vec{q}, \omega) & = &  \chi^{nn}(-\vec{q}, \omega), \chi^{LL}(\vec{q}, \omega)=  \chi^{LL}(-\vec{q}, \omega),
                        \nonumber  \\
  \chi^{nL}(\vec{q}, \omega) & = & - \chi^{Ln}(-\vec{q}, \omega), \chi^{Ln}(\vec{q}, \omega)  =  - \chi^{nL}(-\vec{q}, \omega).
\label{tnl}
\end{eqnarray}
Note that $ \mathcal{P}_x $ and $ \mathcal{T} $ leads to two different equations in the  $nL$ component only.
Combining Eq. (\ref{pxnl}) and Eq. (\ref{tnl}) leads to $ \chi^{nL}(\vec{q}, \omega)=  \chi^{Ln}(\vec{q}, \omega) $.
Directly taking complex conjugate leads to:
\begin{equation}
[\chi^{ij}(\vec{q}, \omega)]^{*}  =  \chi^{ij}(-\vec{q}, -\omega),~~~ i,j=n,L,
\label{ccnl}
\end{equation}
which combining with Eq. (\ref{pxnl}) leads to
\begin{eqnarray}
 Re \chi^{ii}(\vec{q}, \omega) & = & Re \chi^{ii}(\vec{q}, -\omega),~~~ i=n, L,
        \nonumber   \\
 Im \chi^{ii}(\vec{q}, \omega) & = & - Im \chi^{ii}(\vec{q}, -\omega)
\label{cce}
\end{eqnarray}
   and
\begin{eqnarray}
 Re \chi^{nL}(\vec{q}, \omega) & = &  - Re \chi^{nL}(\vec{q}, -\omega),
        \nonumber   \\
 Im \chi^{nL}(\vec{q}, \omega) & = &  Im \chi^{nL}(\vec{q}, -\omega).
\label{ccne}
\end{eqnarray}

 Requiring $\chi^{\mu\nu}(\vec{q},\omega)$ to be Hermite leads to that $\chi^{nn}, \chi^{LL}, \chi^{nL} $ are real. Based on Eq. (\ref{cce}) and (\ref{ccne}), we know that  $\chi^{nn}(\vec{q}, \omega), \chi^{LL}(\vec{q}, \omega) $ are even function of $ \omega $, while $ \chi^{nL}(\vec{q}, \omega) $ is an odd function of $ \omega $. Similarly, $\chi^{nn}, \chi^{LL}, \chi^{nL}$ might also develop imaginary parts: $Im \chi^{ij}(\vec{q}, \omega) \neq 0, i,j=n,L$, which represents the damping of collective modes. In principle, the damping of the collective modes in the ($T_1,T_2$) sector needs not to happen at the same time as that in the $ (n, L) $ sector. However, at the RPA level (see Eq. (\ref{densityspin})), both of them are damped through the PH excitations and therefore they happen at the same time.

These exact relations lead to strong constraints on the behaviors of the collective modes without any approximations. For example, the restored $SU(2)_{spin}$ symmetry at $\vec{q}=0$ dictates that the three gapped modes found here must have the same gap $ \Delta$ at $ \vec{q}=0 $. Then, one can expand the pole equation $\det [\chi^{-1}(\vec{q},\omega)]=0$ near the gap $(\vec{q},\omega)=(\vec{0},\Delta)$ by using the above symmetry constraints on the $\chi^{\mu\nu}$ to obtain the dispersions of the gapped modes. For instance, we find there are two split transverse modes with the dispersion $\omega_{T_{\pm}}=\Delta \pm \beta q+\mathcal{O}(q^2)$ in the $T_1$-$T_2 $ subspace. In the density+longitudinal spin channel, we obtain one gapped longitudinal (called $L$) mode with the dispersion $\omega_L(\vec{q})=\Delta+\alpha q^2 +\mathcal{O}(q^4)$ determined by symmetry. We have compared these qualitative behaviors with the numerical calculation shown in Fig. 1 and demonstrate that the symmetry argument indeed leads to correct physical pictures.

{\sl (3) The symmetry reason why $T_{+} $ and $ T_{-}$ split}

In this part, we show that in order to guarantee the two $ T_{+},T_{-} $ modes to be degenerate, one need to
have both $ \mathcal{T} $ symmetry and the inversion symmetry $ \mathcal{I} $: $ \vec{q} \rightarrow -\vec{q} $.
The $ \mathcal{I} $ symmetry dictates that
\begin{equation}
 \chi^{+-}(\vec{q}, \omega)=  \chi^{+-}(-\vec{q}, \omega), \chi^{-+}(\vec{q}, \omega)=  \chi^{-+}(-\vec{q}, \omega).
\label{i}
\end{equation}
Combining Eq. (\ref{t}) with Eq. (\ref{i}) lead to:
\begin{equation}
 \chi^{+-}(\vec{q}, \omega)= \chi^{-+}(\vec{q}, \omega).
\label{equal}
\end{equation}
Alternatively, if assuming $ \mathcal{S}_x $ symmetry: $ \sigma_x \rightarrow \sigma_x, \sigma_{y,z} \rightarrow -\sigma_{y,z} $, one also find
Eq. (\ref{equal}). The above fact leads to the following conclusion: the SOC term which breaks the inversion symmetry or the $\mathcal{S}_x$ symmetry, thus $\chi^{+-}(\vec{q}, \omega) \neq  \chi^{-+}(\vec{q}, \omega) $ for $ \vec{q} \neq 0 $
which leads to the splitting of the two transverse modes even in the normal phase at any $ \vec{q} \neq 0 $.
Only at $ \vec{q}=0 $, the SOC vanishes and the three modes have the same energy gap.

\section{Collective modes and PH excitations for 3D Fermi gas with Weyl SOC at RPA level}

In the last section, we established exact relations on the density-spin response functions from the general symmetry principle.
Any analytical approximations or numerical calculations must satisfy these exact relations.
In this section, we will perform a concrete calculation on all the response functions at the RPA level whose poles lead to the collective modes.
We also checked that these exact relations due to symmetry are respected in our calculations.

Now we can expand Eq. (M5) $\mathcal{S} [ \phi_{\mu} ]= \mathcal{S}^{(0)}+ \mathcal{S}^{(2)}+ \cdots $ in terms of $\phi_{\mu}$.
As said in the main text, the density channel $\phi_n$ has a non-zero imaginary saddle point value due to the finite particle density of the fermions
which could be eliminated by re-defining $\delta \phi_{\mu}$ as the deviation from its saddle point value $\bar{\phi}_{\mu}$, i.e., $\delta\phi_{\mu}=\phi_{\mu}-\bar{\phi}_{\mu}$. To the second order of $\delta \phi_{\mu}$ (namely RPA level), we obtain
\begin{eqnarray}\label{eq:GaussianHertz}
\mathcal{S}^{(2)}={1\over 2}\int {d^3\vec{k}\over (2\pi)^3}T \sum_n \left({1\over g}\delta^{\mu\nu} - {1\over 4} \bar{\chi}_0^{\mu\nu}(\vec{k},i\omega_n) \right) \delta\phi^{\mu}(\vec{k},i\omega_n)\delta\phi^{\nu}(-\vec{k},-i\omega_n),
\label{second}
\end{eqnarray}
where $\bar{\chi}^{\mu\nu}$ is related to the density-spin susceptibility  $\chi^{\mu\nu}_0(\mathbf{q},i\omega_n)=\langle s^{\mu}(\mathbf{q},i\omega_n)s^{\nu}(-\mathbf{q},-i\omega_n) \rangle$ via $\bar{\chi}^{00}_0=-\chi^{00}_0$, $\bar{\chi}^{0i}_0=i\chi^{0i}_0$, $\bar{\chi}^{i0}_0=i\chi^{i0}_0$, $\bar{\chi}^{ij}_0=\chi^{ij}_0$ with $i,j,=x,y,z$. The non-interacting density-spin susceptibility is given by:
\begin{equation}
\chi^{\mu \nu }_{0}(\vec{q},i\omega_n)=\frac{1}{V}\sum_{k,sr} F_{sr}^{\mu \nu }(k+q,k)\frac{
n_{F}(\xi _{k+q,s})-n_{F}\left( \xi _{k,r}\right) }{i\omega_n -\left( \xi
_{k+q,s}-\xi _{k,r}\right)},
\label{densityspin}
\end{equation}
where  the $ \xi _{k,s}$ is the fermion spectrum with $s$ the helicity $\hat{p}\cdot \vec{\sigma}|ps\rangle=s|ps\rangle$,
the $n_{F}\left( \xi _{k,s}\right) $ is the Fermi distribution function, the $\xi _{k+q,s}-\xi _{k,r}$ is the particle-hole excitation energy and
the overlap factor $F_{sr}^{\mu \nu }(p,q)=\langle ps| \sigma^{\mu}|qr\rangle\langle qr|\sigma^{\nu}|ps\rangle$.
Consistent with the symmetry analysis in Sec. I, the dynamical $ 4 \times 4 $ density-spin susceptibility Eq. (\ref{densityspin}) is split into two $ 2 \times 2 $ subspaces: i) $n-L$ subspace; ii) $T_1-T_2$. The collective modes can be therefore discussed in these two subspaces.

First, we obtain the gapless sound mode $\omega_s=v_s q$ (Fig. 1) in the $n$-$L$ subspace which is determined by the pole of $2 \times 2$ density-spin susceptibility $\det (1+{g\over 4}\sigma_z \chi_{nL})=0$ with
\begin{eqnarray}
\chi^{nL}_{0}(\vec{q},\omega )=
{m(k_{F1}^2+k_{F2}^2)\over 2\pi^2\kappa}
\left(
\begin{array}{cc}
F(y) & \beta y F(y) \\
\beta y F(y) & {1\over 3}+{\sqrt{k_R^2+2m\mu}\over 3k_R}\beta+y^2F(y)
\end{array}
\right),
\label{nl}
\end{eqnarray}
where $k_{F2}>k_{F1}$ are the two Fermi momenta in Fig. 2(a), $2\kappa=k_{F1}+k_{F2}$ ($k_{F2}-k_{F1}$)  for $\mu>0$ ($\mu<0$),
$\beta={k_{F2}^2-k_{F1}^2\over k_{F2}^2+k_{F1}^2}$ and $y=\omega/v_Fq$, $v_F$ the Fermi velocity, $F(y)=1-y \text{arccoth} y $.
It is easy to see that Eq. (\ref{nl}) indeed satisfies the constraints Eqs. (\ref{cce}) and (\ref{ccne}).
We numerically determined the velocity $v_s$ of this gapless sound mode as a function of filling of fermions shown in Fig. 2(b) for three different interaction strengths. This sound mode corresponds to a coupled density-longitudinal spin fluctuation.
There is also a gapped mode in this subspace called L mode shown in Fig. 1.

Then, in the transverse $T_1$-$T_2$ subspace and according to Eq. (\ref{pm}), the two transverse modes are determined at RPA level by
\begin{equation}
1-\frac{g}{2}\chi^{+-}_{0}(\vec{q}, \omega)=0,~~1-\frac{g}{2} \chi^{-+}_0(\vec{q}, \omega)=0.
\label{poles}
\end{equation}
where $ \chi^{+-}_0(\vec{q}, \omega)= 2( \chi^{xx}-i \chi^{xy} ) $ is given by Eq. (\ref{densityspin}). Numerically solving this pole equation leads to two transverse spiral modes shown in Fig. 1.

\begin{figure}[!t]
\begin{center}
\includegraphics[width=6in]{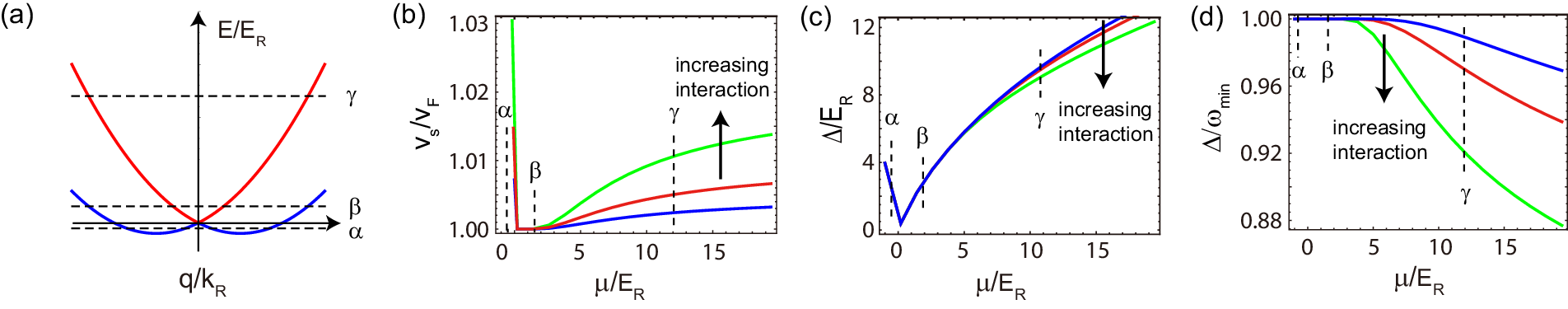}\hspace{0.5cm}
\end{center}
\caption{(Color online) (a) Single particle energy spectrum of SOC Fermi gas in the normal state. Three typical filling cases are denoted by the dashed lines $\alpha$, $\beta$ and $\gamma$, which corresponds to strong, intermediate and weak SOC, respectively. (b) shows the ratio $v_s/v_F $ of gapless sound mode as a function of filling. (c) $ \Delta/E_R $ and (d) $  \Delta/\omega_{min} $ show the energy gap for the three gapped modes as a function of filling. The three lines with  different colors represent different interaction strengths, i.e., green (critical interaction strength $g_c$ for ferromagnetic phase transition), red ($g=4/5g_c$), blue ($g=2/3g_c$).}
\label{fig:coll}
\end{figure}

Due to the $SU(2)_{spin}$ symmetry at $\vec{q}=0$, the three gapped modes are degenerate at $\vec{q}=0$ where the gap is
determined by $\det[1-{g\over 4} \chi^{ij}_{spin}(\omega,\vec{q}=0)]=0$, with
\begin{eqnarray}
\chi^{ij}_{spin}(\omega,\vec{q}=0) = {m\over 6\pi^2k_R}\left( k_2^2-k_1^2+{m^2\omega^2\over 4k_R^2} \ln \left[ {4k_R^2k_2^2-m^2\omega^2\over 4 k_R^2k_1^2-m^2\omega^2} \right] \right)\delta^{ij}, i,j=x,y,z,
\end{eqnarray}
The numerical result of this energy gap $ \Delta $ at different $ \mu/E_R $ is shown in Fig. 2(c) (in units of the recoil energy $E_R=k_R^2/2m$)
and Fig. 2(d) (in units of the lower edge of inter-band particle-hole excitation energy $\omega_{min}$ shown in Fig. 1).

In Fig. 2(a), we plot three typical filling cases shown by the dashed lines: $\alpha$) dilute density case with $\mu<0$; $\beta$) $\mu>0$ but close to the Dirac point ($\mu=0$), $\gamma$) dense density case. Fig. 1 belongs to the case $\gamma$).
In Fig. 2(b), as the filling tends to the Dirac point from above $\mu>0$, the $ v_s/v_F $ decreases close to 1.
 Passing the Dirac point, the ratio increases quickly. In Fig. 2(c), the $\Delta/E_R $ decreases close to zero when approaching the Dirac point, where the blank regime between the inter-band and intra-band PH excitation shown in Fig. 1 shrinks to zero.
 Passing the Dirac point, the ratio increases and the blank regime between the inter-band and intra-band PH excitation also expands.
  To show the relationship between the energy gap $\Delta$ and the PH excitation, in Fig. 2(d) we show $\Delta/\omega_{min}$ with $\omega_{min}$
  the lower edge of the inter-band PH excitation energy shown in Fig. 1.
  We find as the chemical potential decreases from case $\gamma$) to $\alpha$), the ratios gets to 1 from below, namely, the gapped collective modes move
  more closer to the edge of inter-band particle-hole excitation continuum in Fig. 1.
  In the dilute limit $\mu\rightarrow -1$, the critical interaction strength to get to ferromagnetic state
  $g_c=12\pi^2\sqrt{1+\mu}+\mathcal{O}((1+\mu)^{3/2})$ reduces to zero, the sound velocity saturates $v_s/v_F \simeq 1.102 $,
  and the energy gap increases and moves close to the PH continuum $\Delta_s/E_R \simeq 4(1-\sqrt{1+\mu})$.

  The level repulsion in the $ n$-$L $ space will push the sound mode quite close to the intra-band PH continuum (Fig. 1,2(b)).
  Similarly, the level repulsion in the $ T_1, T_2 $ space will also split the two transverse modes $ T_{\pm} $.
  So at finite, but small $ q $, the three modes split with  $ \omega_{T_{+} } > \omega_{L} > \omega_{T_{-}} $.